\documentclass[journal]{IEEEtran}
\IEEEoverridecommandlockouts
\usepackage{cite}
\usepackage{amsmath,amssymb,amsfonts}
\usepackage{algorithmic}
\usepackage{graphicx}
\usepackage{subfigure}
\usepackage{textcomp}
\usepackage{xcolor}
\usepackage{gensymb}
\usepackage{multirow}
\usepackage{booktabs}  
\usepackage{threeparttable}
\usepackage{epstopdf}
\usepackage{amssymb}
\usepackage{makecell}
\usepackage{upgreek}
\usepackage{bm}
\usepackage{multirow}
\usepackage{makecell}
\usepackage{url}
\usepackage{dblfloatfix}
\makeatletter
\renewcommand{\maketag@@@}[1]{\hbox{\m@th\normalsize\normalfont#1}}%
\makeatother

\def\BibTeX{{\rm B\kern-.05em{\sc i\kern-.025em b}\kern-.08em
    T\kern-.1667em\lower.7ex\hbox{E}\kern-.125emX}}
\begin{document}
\title{3GPP-Like GBSM THz Channel Characterization, Modeling, and Simulation Based on Experimental Observations
\thanks{Manuscript received xxx; revised xxx; accepted xxx. Date of publication xxx; data of current version xxx. This work was supported in part by the National Science Fund for Distinguished Young Scholars (No. 61925102),  the National Natural Science Foundation of China (Nos. 62201086, 62101069, and 62201087), and BUPT-CMCC Joint Innovation Center. $\textit{(Corresponding \ auther: \ Jianhua \ Zhang.)}$\\
Zhaowei Chang, Jianhua Zhang, Pan Tang, Lei Tian, and Ximan Liu are with the State Key Lab of Networking and Switching Technology, Beijing University of Posts and Telecommunications, Beijing 100876, China (e-mail: \{changzw12345, jhzhang, tangpan27, tianlbupt, liuxm2020\}@bupt.edu.cn).\\
Hao Jiang is with the School of Artificial Intelligence/School of Future Technology, Nanjing University of Information Science and Technology, Nanjing 210044, China; and also with the National Mobile Communications Research Laboratory, Southeast University, Nanjing 210096, China (e-mail: jianghao@nuist.edu.cn).\\
Guangyi Liu is with China Mobile Research Institute, Beijing 100053, China (e-mail: liuguangyi@chinamobile.com).
}
}

\author{\IEEEauthorblockN{Zhaowei Chang, Jianhua Zhang, Pan Tang, Lei Tian, Hao Jiang, Ximan Liu, and Guangyi Liu
}\\
}
\maketitle

\begin{abstract}
Terahertz (THz) communication is envisioned as one of the possible technologies for the sixth-generation (6G) communication system due to its rich spectrum. To evaluate the performance of THz communication, it is essential to propose THz channel models within the common framework of the geometry-based stochastic model (GBSM) in the 3rd Generation Partnership Project (3GPP). This paper focuses on THz channel modeling and simulation by a 3GPP-like GBSM, based on channel measurements. We first present channel measurements at 100 GHz in an indoor office scenario and 132 GHz in an urban microcellular scenario. Subsequently, channel characteristics such as path loss, delay spread, angle spread, K-factor, cluster characteristic, cross-correlations, and correlation distances are obtained and analyzed based on channel measurement. Additionally, the channel characteristics are modeled by the statistical distribution of 3GPP channel models, which can be used to reconstruct the channel impulse response (CIR). Furthermore, these obtained distributions are studied referring to the default models in the 3GPP, revealing the channel sparsity in the THz channel. For instance, in the case of line-of-sight links in the indoor office, the mean of the measured cluster number is 4 while the default value is 15. Finally, we propose the THz channel model and its simulation framework to reconstruct CIRs based on the obtained models, which aim at characterizing the sparser THz channels. The obvious channel sparsity is characterized in both scenarios, as the Gini factors obtained by the proposed model only have the maximum deviation of 0.04 for those of the measurement. Overall, these findings are helpful in understanding and modeling the THz channel, facilitating the application of THz communication techniques for 6G.
\end{abstract}

\begin{IEEEkeywords}
Terahertz, channel measurement, 3GPP-like channel model, GBSM, THz communication system 
\end{IEEEkeywords}

\section{INTRODUCTION}\label{1}
\subsection{Background}
\IEEEPARstart{A} very wide band is needed to achieve the ultra-reliable transmission to support various critical applications, such as cloud computing and space-air-ground integrated networks, following the development of big data \cite{r444}, \cite{r30}, \cite{r5}. The International Telecommunication Union (ITU) has already estimated that the number of global mobile subscriptions could reach 13.8 billion in 2025 and 17.1 billion in 2030 \cite{r22}, \cite{ttst}, \cite{r2}. Therefore, Terahertz (THz) communication attracts a great deal of attention due to its wide bandwidth between 0.1 to 10 THz \cite{r7}, \cite{r3}, which supports high-speed data rates from tens of Gbps to several Tbps \cite{r32}, \cite{r8}. Thus, the THz spectrum would be considered a reliable frequency band for the sixth-generation (6G) communication system \cite{r31}, \cite{yuan}. 

As the frequency goes up to THz and the wavelength gets smaller, the short wavelength of the THz bands would close to particles in the environment, diffraction is weaker and reflection gradually occupies a dominant situation. As the THz spectrum has changed compared to millimeter wave (mm-Wave), the fundamental challenges of designing, evaluating, and optimizing THz wireless communication systems include knowing the laws of electromagnetic propagation, thorough investigation of channel characteristics, and construction of efficient and accurate channel model in the THz bands \cite{hanchong}. For instance, the THz path loss (PL) models can be used in link budget calculations and cellular coverage predictions \cite{rappl}, and the delay of the radio propagation acts on the modulation scheme such as the guard interval of orthogonal frequency division multiplexing (OFDM) set to commonly four times of root-mean-square (RMS) delay guard interval. To achieve these goals, one of the main approaches, physical channel measurement, could acquire the real THz propagation characteristics \cite{thztds}. Thus, in this paper, we carry out the realistic channel measurements in the indoor office and the urban microcellular (UMi) scenario, give a comprehensive analysis and modeling of channel characteristics, and propose the channel model and modeling implementation framework in the THz bands.
\subsection{Related Works}
A number of measurement campaigns have been conducted in the THz bands recently. The measurement scenarios are mainly divided into short-range scenarios, typical indoor scenarios, and outdoor scenarios. For the short-range scenarios, the measurement is conducted at a straight path within the distance of 5.5 $\rm{m}$ at 140-150, 180-190, and 210-215 $\rm{GHz}$ \cite{r13}. The path loss exponents (PLE) of around 2.1-2.2 are obtained. Similarly, the authors in \cite{r14} carried the measurement within 4 $\rm{m}$ at 110, 140, and 170 $\rm{GHz}$ and got the PLE of 1.8-1.9. Besides, the measurements in the small open room at 140, 150, 285 $\rm{GHz}$ \cite{r16} to obtain the PLE, and the value of delay spread (DS), angular spread. For the typical indoor scenarios, the office room, corridor, and lobby are the main measurement and modeling scenarios. For example, the measurement in the office room are conducted at 142 $\rm{GHz}$ \cite{140office}. Based on the measurement, the PLE, DS, and the number of clusters and paths are compared between 28, 73, and 142 GHz. In \cite{r19}, more channel characteristics are investigated, i.e., the path loss (PL), DS, angular spread, cluster characteristics, and cross-correlations based on the office room measurement at 140 $\rm{GHz}$. The corridor and lobby are measured to extract the value of DS and PL at 300 $\rm{GHz}$ in \cite{r20}. For the outdoor scenarios, for example, the measurement in the street canyon at 145 to 146 $\rm{GHz}$ is conducted to study the parameters of the model for the PL, shadowing, DS, angular characteristic, and multipath component (MPC) power distribution \cite{r34}. The channel measurements in UMi scenarios at 159 $\rm{GHz}$ \cite{r35} and 142 $\rm{GHz}$ \cite{r36} are conducted to study the PL, delay-domain, and angle-domain characteristics. In short, the research on measurements of the THz channels is mainly around 140, 150, and 300 $\rm{GHz}$ in typical scenarios. The study of parameters analysis and modeling contains PL, DS, and angular characteristics, and the categories of characteristic statistical modelings are not sufficient for THz channel modeling.
\subsection{Motivation}
The approval of the 6G framework and overall objectives by the ITU Radiocommunication Working Party 5D in June 2023 \cite{r39} signifies the official commencement of 6G standardization. Consequently, to facilitate the assessment of THz technology, the development of a geometry-based stochastic model (GBSM) THz channel model, similar to the 3rd Generation Partnership Project (3GPP), is imperative. However, existing studies on the THz channels have predominantly focused on its partial characteristics investigation, which did not conduct sufficient descriptions of channel characteristics for catering to the modeling requirements. For example, GBSM, which mainly operates in the case of recognizing the distribution of large-scale parameters, pairwise correlations in the spatial domain, and cluster distribution. Additionally, the THz frequency bands allocated for mobile usage by the ITU, such as the bands below 140 $\rm{GHz}$ (130-134 $\rm{GHz}$), have not been extensively investigated. These factors have motivated us to conduct measurements, characteristics analysis, and modeling to develop the THz channel modeling. This research aims to bridge the gaps in terms of frequency bands, scenarios, and parameter modeling of the THz channel, with the ultimate goal of enabling the design, deployment, evaluation, and optimization of THz communication systems.   
\subsection{Main Contributions}
In this paper, we present two measurement campaigns in typical hotspot scenarios, specifically at 100 $\rm{GHz}$ in an indoor office and 132 $\rm{GHz}$ in UMi scenarios. These scenarios include both line-of-sight (LoS) and non-line-of-sight (NLoS) conditions. Based on the measurements, we extract and analyze various parameters such as PL, DS, azimuth angle spread of arrival (ASA), K-factor, cluster number, cluster DS, cluster ASA, cluster K-factor, cross-correlations, and correlation distance. The channel characteristics are then statistically modeled based on the statistical distribution of the 3GPP channel model. We analyze the measured statistical models referring to the default models in the 3GPP and observe the distinct channel sparsity from the analysis. Finally, we propose the THz channel model and its implementation framework for simulation to reconstruct channel impulse responses (CIRs) by the statistical models of the channel characteristic parameters from the measurements. Through the framework, the sparse THz channel is described, which is verified by the items of CIRs, Gini factors, and the channel capacity in the simulated THz channel. The contributions of this paper are as follows:
\begin{itemize}
\item We conduct two extensive channel measurement campaigns, they are in the indoor office scenario at 100 $\rm{GHz}$ and the UMi scenario at 132 $\rm{GHz}$. For the measurement at 100 $\rm{GHz}$, a total of 19080 (54 rotation angles $\times$ 18 measurement points $\times$ 20 snapshots) CIRs are collected. For the measurement at 132 $\rm{GHz}$, the total 907200 (108 RX rotation angles $\times$ 7 TX rotation angles $\times$ 24 measurement points $\times$ 50 snapshots) CIRs are collected. 
\item Based on the measurement, we extract and analyze the power delay profile (PDP), and channel characteristics, i.e., the PL, DS, ASA, K-factor, cluster number, cluster DS, cluster ASA, cluster K-factor, cross-correlations, and correlation distance. Then, representative measurement points are resolved by path-tracing derivation, and the statistical models of the channel characteristics are proposed according to the statistical distribution models of the 3GPP channel model. These obtained values can be used in the channel model standardization. 
\item For the sake of investigating the propagation difference in the channel between THz and low frequency and between scenes, we compare the channel statistical characteristics for different scenarios (LoS-NLoS and the indoor office-the UMi), analyze those characteristics by referring to the 3GPP channel model available less than 100 $\rm{GHz}$ at the same scenario. The obvious channel sparsity of the THz channel is observed by the investigation, which enlightens an aspect of the THz channel modeling.
\item The THz channel model and its simulation implementation framework are proposed to obtain the CIRs based on the GBSM, which utilizes the measured statistical models of channel characteristic parameters. By calculating the CIRs, Gini factors, and channel capacity, it is observed that the THz channel model and its simulation framework successfully characterize the channel sparsity in the THz channel.
\end{itemize}

The remainder of the paper is organized as follows. Section \ref{2} provides an introduction to the setup of the channel sounder, the measurement environment, and the measurement procedures. Section \ref{3} covers the data processing, and calculation of channel characteristics. Section \ref{4} analyzes and models the channel characteristics, and studies the measured results by applying the default models in the 3GPP. Section \ref{5} presents the THz model and its implementation framework, and simulates the CIRs, Gini factors, and channel capacity to evaluate the performance of the THz channel model. Finally, Section \ref{6} presents the conclusion of this paper.
\section{TERAHERTZ CHANNEL MEASUREMENTS}\label{2}
The channel model forms the cornerstone of the technology research and system trial \cite{r21}. Conducting wireless propagation channel measurements \cite{r1} is the most direct and effective way to obtain these channel models, as measurements can account for greater variability in environments compared to simulation. In this paper, channel measurements are conducted using a channel sounder for the indoor office scenario at 100 $\rm{GHz}$, which represents the upper boundary of the current 3GPP working frequency band. Additionally, measurements are conducted for the UMi scenario at 132 $\rm{GHz}$, which are the unexplored frequency points within the bands (130-134 GHz) allocated by the ITU for mobile use \cite{r38}. The measurement setup, procedures, and data processing will be introduced in the following section. The measurement setup, procedures, and data processing will be introduced in the following part.
\begin{table}[t]
\begin{center}
\setlength{\tabcolsep}{7pt}  
\renewcommand\arraystretch{1.3}  
\caption{THE CHANNEL SOUNDER SETUP.}\label{Table 1}
\vglue8pt
\begin{tabular}{ccc}  
 \hline
  {\textbf{Parameter}}   &{\textbf{Value of Office}} &{\textbf{Value of UMi}}\\     
 \hline
   {Tx height}  &{1.8 $\rm{m}$} &{11.6 $\rm{m}$}\\
    \hline
   {Rx height}  &{1.5 $\rm{m}$} &{1.5 $\rm{m}$}\\
    \hline
   {Centre frequency}  &{100 $\rm{GHz}$} &{132 $\rm{GHz}$}\\
    \hline
   {Bandwidth}  &{1.2 $\rm{GHz}$}  &{1.2 $\rm{GHz}$}\\
    \hline
   {Sampling rate}  &{1.5 $\rm{GHz}$} &{1.2 $\rm{GHz}$}\\
    \hline
   {Sequence samples}  &{511} &{511}\\
    \hline
   {Sequence duration}  &{0.85 $\upmu$s}  &{0.85 $\upmu$s}\\
    \hline
   {IF frequency}  &{4 $\rm{GHz}$} &{6 $\rm{GHz}$}\\
    \hline
   {LO frequency}  &{12 $\rm{GHz}$}  &{21 $\rm{GHz}$}\\
    \hline
   {Multiplying factor}  &{8} &{6}\\
    \hline
   {Antenna gain at TX}  &{8 $\rm{dBi}$}  &{22.5 $\rm{dBi}$}\\
    \hline
   {Antenna gain at RX}  &{19.7 $\rm{dBi}$}  &{24.5 $\rm{dBi}$}\\
    \hline
   {Horizontal/vertical HPBW at TX}  &{60$^\circ$/60$^\circ$} &{14$^\circ$/12$^\circ$}\\
    \hline
    {Horizontal/vertical HPBW at RX}  &{20$^\circ$/15$^\circ$}  &{10$^\circ$/9$^\circ$}\\
    \hline
   {Antenna polarization at TX}  &{Circular}  &{Vertical}\\
    \hline
   {Antenna polarization at RX}  &{Vertical}  &{Vertical}\\
    \hline
   {Azimuth rotation range at RX}  &{[0$\degree$, 360$\degree$)} &{[0$\degree$, 360$\degree$)}\\
    \hline
   {Elevation rotation range at RX}  &{[-15$\degree$, 15$\degree$]} &{[-9$\degree$, 9$\degree$]}\\
    \hline
   {Azimuth rotation step}  &{20$\degree$} &{10$\degree$}\\
    \hline
   {Elevation rotation step}  &{15$\degree$} &{9$\degree$}\\
    \hline
\end{tabular}
\end{center}
\end{table}
\subsection{Channel sounder setup}\label{21}
\begin{figure}[!t]
\centerline{\includegraphics[scale=0.3]{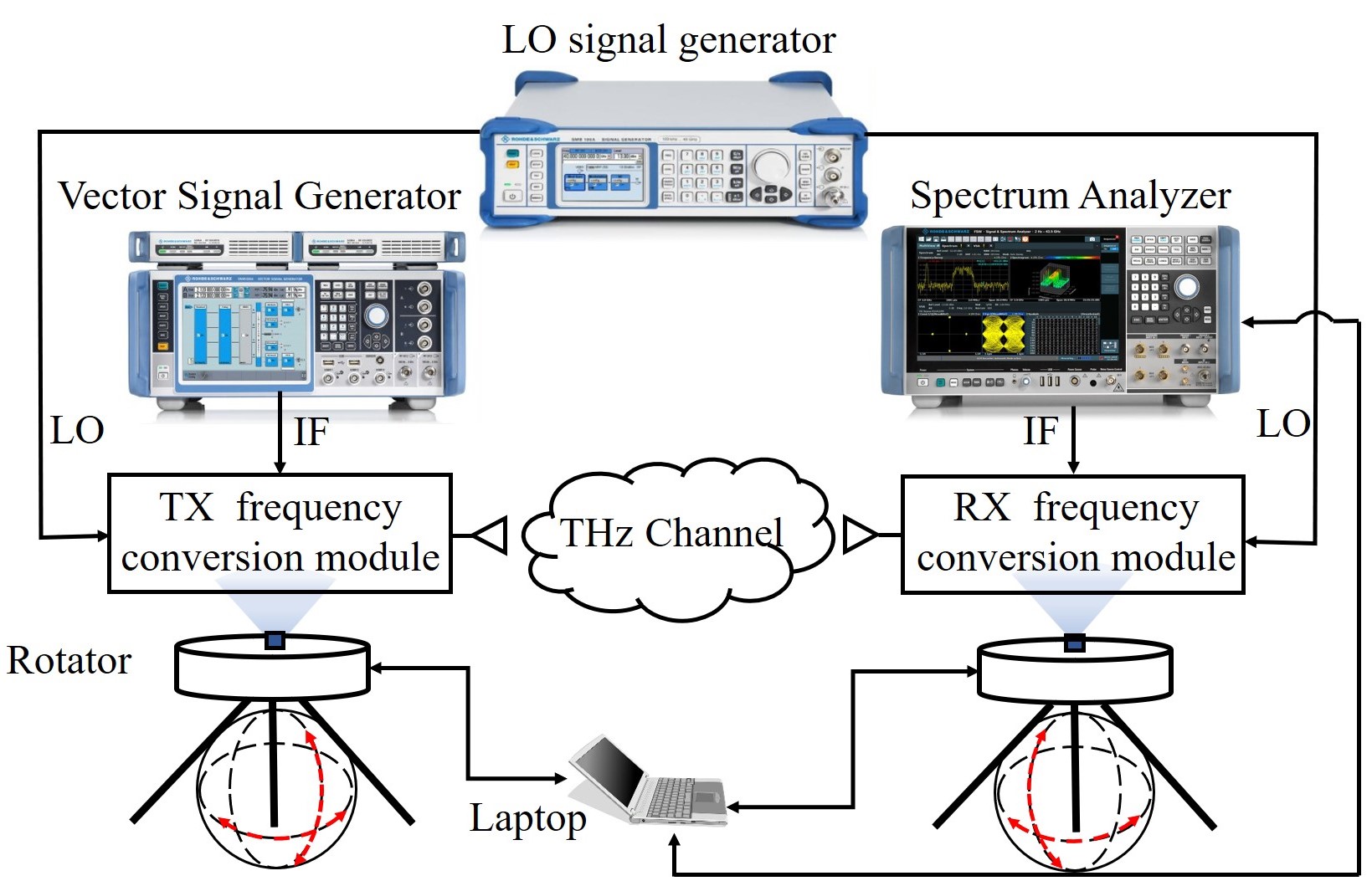}}
\caption{The structure of channel sounding platform.}\label{Fig 2}
\end{figure}

At the transmitter (TX) side, a vector signal generator (R\&S SMW 200A) generates the intermediate frequency (IF) signal. The signal is a modulated signal using amplitude shift keying (ASK) from a periodic 511 samples Pseudo-Noise 9 (PN9) sequence with a duration of 0.85 $\upmu$s. The signal propagates through the cable and multiplies a local oscillator (LO) signal generated by the signal generator (R\&S SMB 100A) using the frequency multiplier. Thus, the IF signal is mixed with the LO signal and multiplied to the THz bands. A horn antenna with high gain is used to transfer the THz signal. At the receiver (RX) side, a horn antenna is used to receive the THz signal. Then, the THz signal is down-converted by a frequency mixer and becomes the IF signal. A spectrum analyzer (R\&S FSW 43) catches the IF signal and demodulates it. The CIRs are obtained after the data processing by the laptop. The channel platform is shown in Fig. \ref{Fig 2}.

For the measurement at 100 GHz, the frequency multiplying factor is 8. The IF and LO frequencies are 4 and 12 $\rm{GHz}$, respectively. At the RX side, 1278 in-phase and quadrature (IQ) signal samples are obtained with a sample rate of 1.5 $\rm{GHz}$. TX and RX antenna gains are 8 and 19.7 $\rm{dBi}$ respectively. Half power beam width (HPBW) at TX and RX in the horizontal plane are 60$\degree$ and 20$\degree$, and in the vertical plane is 60$\degree$ and 15$\degree$, respectively. The circular polarization antenna at TX provides wide beamwidth but reduces about $3$ $\rm{dBi}$ gain due to the vertical polarization reception. 

For the measurement at 132 GHz, the frequency multiplying factor of the frequency multiplier is 6. The IF and LO frequencies are 6 and 21 $\rm{GHz}$, respectively. Besides, 1022 IQ signal samples are obtained with a sample rate of 1.2 $\rm{GHz}$. The lower sample rate is used in the large scenario to reduce the sampling time under the premise of sufficient delay resolution. The antenna gains at TX and RX are 22.5 and 24.5 $\rm{dBi}$, respectively. The HPBWs of the Tx antenna are 14$\degree$ and 12$\degree$ for horizontal and elevation planes. The HPBWs of RX antenna for horizontal and elevation plane are 10$\degree$ and 9$\degree$. The details of the channel sounder configuration for the measurement campaigns are shown in Table \ref{Table 1}.
\subsection{Measurement environment and procedures}\label{22}
\begin{figure}[!t]
\centering
\subfigure[The layout in the indoor office]{\includegraphics[width=4.35cm]{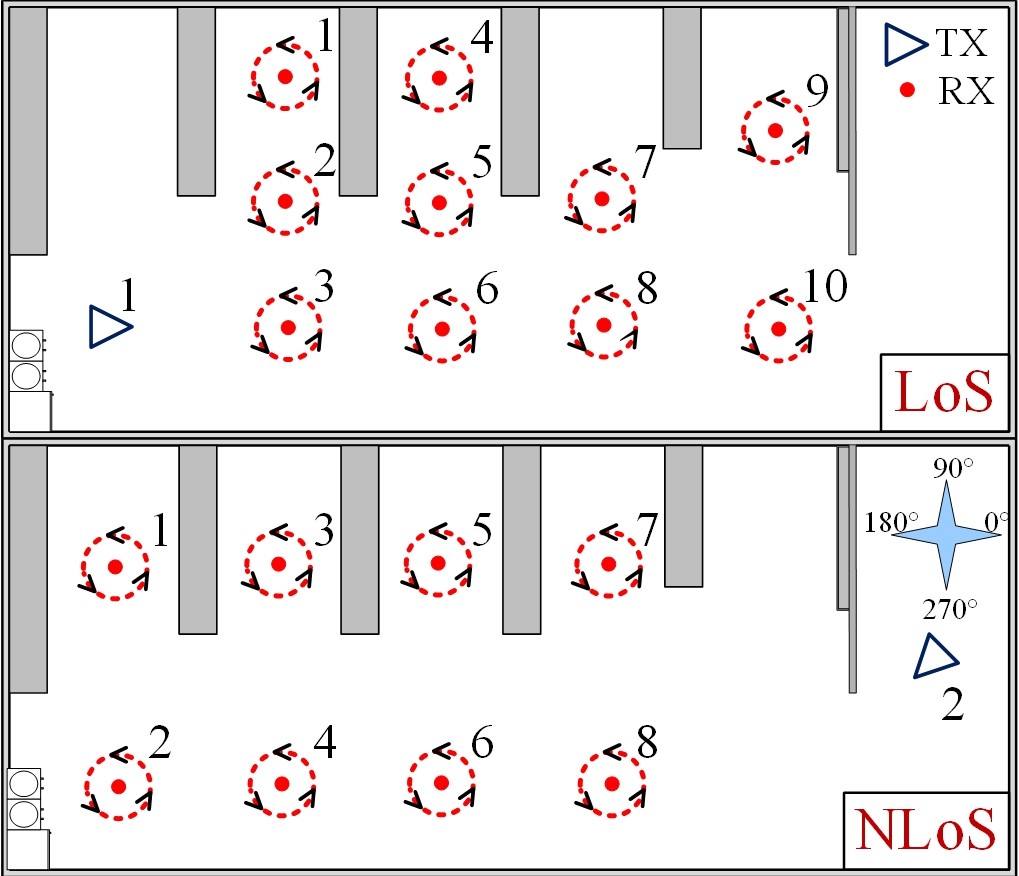}}
\subfigure[The measurement environment]{\includegraphics[width=4.35cm]{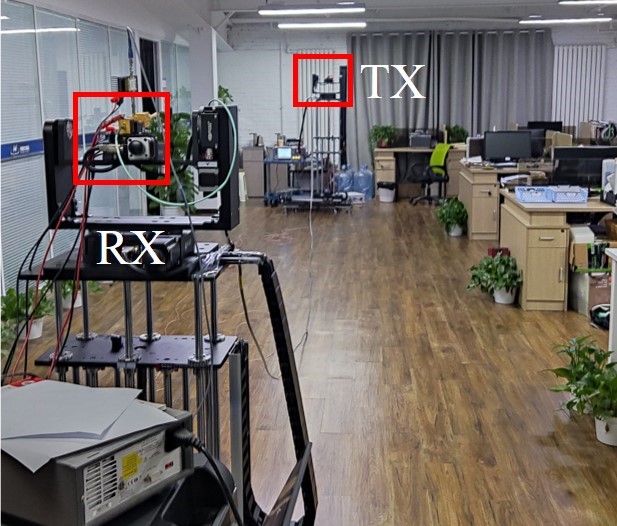}}
\caption{The measurement layout (a) and the measurement environment (b) in the office office.}\label{Fig 3}
\end{figure}

For the measurement at 100 GHz, the channel measurement is conducted in an indoor office room with an area of 17.51$\times$6.78 $\rm{m^2}$. The office contains an open cubicle area, a walled office, an open area, and a corridor, etc. Fig. \ref{Fig 3}(a) illustrates the measurement layout for both LoS and NLoS cases. The measurement environment in the indoor office is shown in Fig. \ref{Fig 3}(b). The top of Fig. \ref{Fig 3}(a) displays the locations of the TX and RX in the LoS case. The LoS locations are distributed in the open cubicle areas and open areas. The RXs in the LoS case are also affected by the walled office between the $\rm{TX1_I}$. Besides, $\rm{TX2_I}$ in the corridor radiates to the office area to investigate the case that TX/RX is interfered with by the corridor walls. The distribution of TX and RX is representative of the indoor office scenario and covers the definition of the 3GPP. In detail, $\rm{TX1_I}$ serves ten RX locations, namely $\rm{RX1_{IL}}$ to $\rm{RX10_{IL}}$ with corresponding distances of 4.5, 3.2, 2.7, 6.3, 5.7, 5.5, 8.5, 8.3, 10.4, and 10.1 $\rm{m}$, respectively. For each TX-RX in the LoS case, one TX best-pointing elevation angle that maximizes the received power at RX is set and three elevation angles are used at RX, i.e., the RX best-pointing elevation angle, and the RX antenna uptilts and down-tilts by 15$\degree$ from the RX best-pointing elevation angle. For each TX and RX elevation angle combination, the TX best-pointing azimuth angle is set and the RX antenna is rotated 360$\degree$ in the azimuth plane with a step of 20$\degree$ to catch all the possible multi-paths in any azimuth directions. All steps of the rotation are selected to one HPBW of the corresponding antennas. Below Fig. \ref{Fig 3}(a) are the measurement locations in the NLoS case. $\rm{TX2_I}$ serves eight RX locations, namely $\rm{RX1_{IN}}$ to $\rm{RX8_{IN}}$, with distances of 14.2, 14.0, 11.2, 10.9, 8.6, 8.1, 5.5, and 4.2 $\rm{m}$, respectively. For each TX-RX in the NLoS case, the horizontal angle is used at TX and three elevation angles are used at RX, i.e., the horizontal angle, and the RX antennas uptilt and down-tilt by 15$\degree$ from the horizontal angle. TX radiates towards the wall at a 45$\degree$ incident angle in the azimuth plane. The RX azimuth plane rotation in the NLoS case follows the same approach as in the LoS case. In the measurement for each TX-RX elevation and azimuth angle combination, 20 snapshots of IQ data were collected.
\begin{figure}[!t]
\centering
\subfigure[The layout in the UMi]{\includegraphics[width=4.35cm]{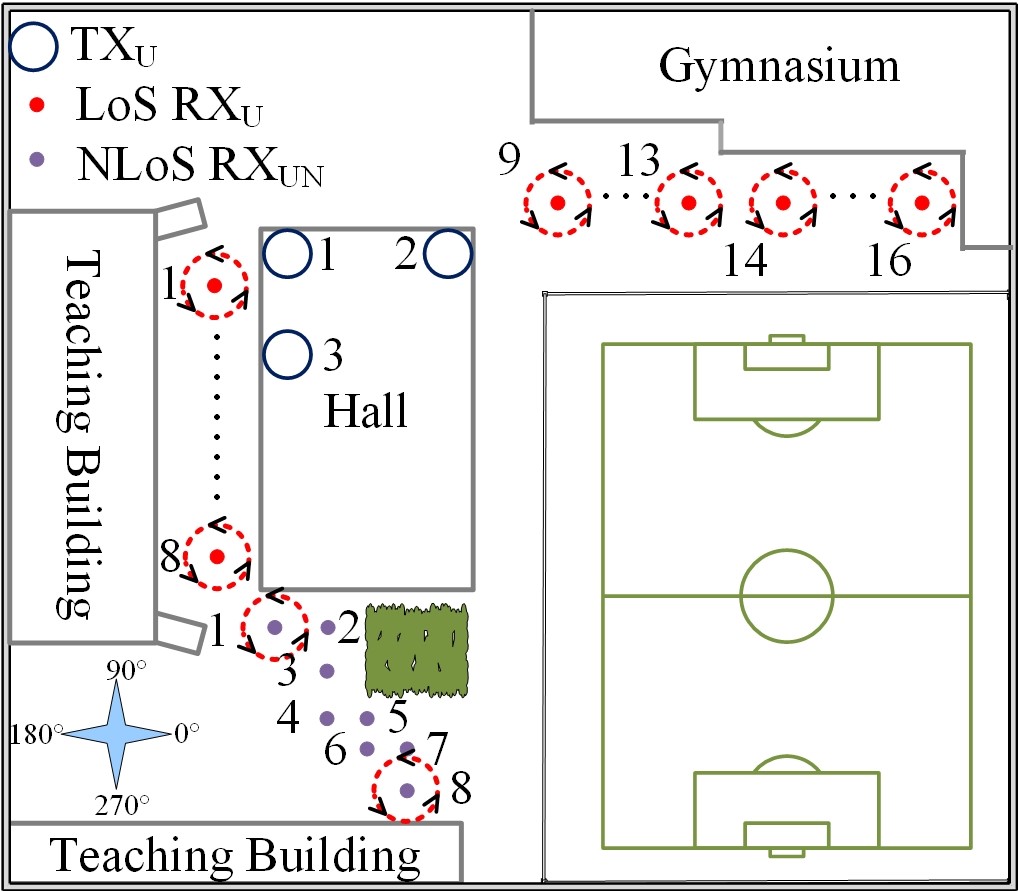}}
\subfigure[The measurement environment]{\includegraphics[width=4.35cm]{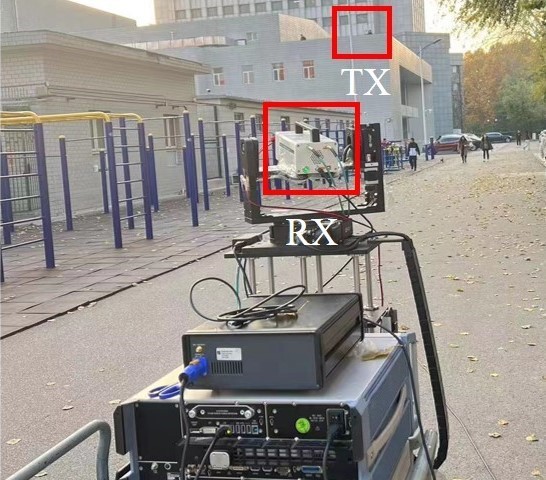}}
\caption{The measurement layout (a) and the measurement environment (b) in the UMi.}\label{Fig 4}
\end{figure}
For the measurement at 132 GHz, the UMi scenario and locations of measurement points of channel measurement are illustrated in Fig. \ref{Fig 4}(a). The measurement environment in the UMi is shown in Fig. \ref{Fig 4}(b). TX is set at $\rm{TX1_U}$ and $\rm{TX2_U}$ for LoS case and set at $\rm{TX3_U}$ for NLoS case. The TX and RX heights are maintained at 11.6 and 1.5 $\rm{m}$, respectively, throughout the measurement. Also, 24 locations of RX ($\rm{RX1_{UL}}$-$\rm{RX16_{UL}}$) in the LoS case and RX ($\rm{RX1_{UN}}$-$\rm{RX8_{UN}}$) are set at a total of three routes as shown in Fig. \ref{Fig 3}(b). $\rm{RX1_{UL}}$-$\rm{RX8_{UL}}$ are set to investigate the street canyon area, and $\rm{RX9_{UL}}$-$\rm{RX16_{UL}}$ are represent the urban open area. $\rm{RX1_{UN}}$-$\rm{RX8_{UN}}$ are blocked by the body of the hall, and can be utilized to study the NLoS case. The set of TX and RX is typical for the UMi scenario. The distances between $\rm{TX1_U}$ and $\rm{RX_{UL}}$ for LoS measurements are as follows: $\rm{RX1_{UL}}$ - 14.2 $\rm{m}$, $\rm{RX2_{UL}}$ - 17.5 $\rm{m}$, $\rm{RX3_{UL}}$ - 21.6 $\rm{m}$, $\rm{RX4_{UL}}$ - 25.9 $\rm{m}$, $\rm{RX5_{UL}}$ - 30.5 $\rm{m}$, $\rm{RX6_{UL}}$ - 35.1 $\rm{m}$, $\rm{RX7_{UL}}$ - 39.9 $\rm{m}$, and $\rm{RX8_{UL}}$ - 44.7 $\rm{m}$. Besides, for $\rm{TX2_U}$ and $\rm{RX_{UL}}$, the distances from $\rm{RX9_{UL}}$ to $\rm{RX16_{UL}}$ in the LoS case are 53.9, 58.7, 63.5, 68.4, 73.3, 83.0, 92.8, and 102.7 $\rm{m}$, respectively. For each $\rm{RX_{UL}}$ location in the LoS case, $\rm{TX_{U}}$ is aligned with RX at the elevation angle, and seven azimuth angles are used at $\rm{TX_{U}}$ (the aligned angle and rotations of the $\rm{TX_U}$ antenna by 14 degrees in both clockwise and counterclockwise directions, repeated three times). The distances between $\rm{TX3_U}$ and $\rm{RX_{UN}}$ from $\rm{RX1_{UN}}$ to $\rm{RX8_{UN}}$ in the NLoS case are 56.7, 56.9, 61.8, 67.7, 68.1, 71.4, 71.9, and 76.4 $\rm{m}$, respectively. For each $\rm{RX_{UN}}$ location in the NLoS case, $\rm{TX_U}$ and $\rm{RX_{UN}}$ are in the direction of maximum primary reflection. For each TX-RX combination, three elevation angles are utilized at RX (the aligned angle, and the RX antennas uptilt and down-tilt 9$\degree$ from the aligned angle), and the RX antenna is rotated 360$\degree$ in the azimuth plane with a step of 10$\degree$ to catch all the possible rays in any azimuth directions. In the measurement for each TX-RX elevation and azimuth angle combination, 50 snapshots of IQ data were collected.
\section{Processing of Measurement}\label{3}
After conducting the measurement described in Section \ref{2}, the data needs to be processed to generate the power delay profile (PDP). The calibration procedure for the system is explained in detail in \cite{r21}. This procedure involves removing the system and antenna response from the original data through calibration. Subsequently, the omnidirectional PDP and other channel characteristics can be derived from the directional PDPs. This section will cover the synthesis of PDPs and the computation of channel characteristics.
\subsection{Synthesizing PDPs}\label{31}
The obtained CIRs after the system calibration are directional, representing the channel responses for specific azimuth and elevation angles. The thresholds are selected to be 10 $\rm{dB}$ above the average noise power (ANP) of PDPs in the indoor office scenario and 6 $\rm{dB}$ above the ANP of PDPs in the UMi scenario (low noise amplifier with better performance is used for the UMi measurement) to reduce the impact of noise in the results. In order to get the omnidirectional PDP, we synthesize the directional PDPs by selecting the strongest path among all azimuth and elevation directions for each delay bin:
\begin{equation}\label{S}
\begin{split}
h_\text{omni}(\tau)=\max\limits_{\phi_\text{TX},\phi_\text{RX},\theta_\text{RX}}h(\tau,\phi_\text{TX},\phi_\text{RX},\theta_\text{RX}),
\end{split}
\end{equation}
\noindent where $h$ denotes the directional PDPs, $\phi_\text{TX}$ and $\phi_\text{RX}$ are the azimuth orientation of TX and RX, respectively, $\theta_\text{RX}$ is the elevation orientation of RX, and $\tau$ is per delay bin.
\subsection{Computing channel characteristics}\label{33}
To characterize the THz propagation channel, the computation of the channel characteristics based on the directional and omnidirectional PDP is introduced in this part.
\subsubsection{Path loss and shadowing}\label{331}
PL reflects the influence of the distance between the TX and RX on the quality of the received signal \cite{jh}. The omnidirectional and the strongest directional PL is calculated by the negative value of the sum of the power of all the paths on each delay bin \cite{moliTHz}:
\begin{equation}\label{O}
\begin{split}
PL_\text{O}[{\rm{dB}}]=-\sum_{i=1}^L\left|h_\text{omni}(\tau_i)\right|^2,
\end{split}
\end{equation}
\begin{equation}\label{B}
\begin{split}
PL_\text{B}[{\rm{dB}}]=-\sum_{i=1}^L\left|h(\tau_i,\Phi_\text{RX},\Theta_\text{RX})\right|^2,
\end{split}
\end{equation}
\noindent with
\begin{equation}\label{J}
\begin{split}
(\Phi_\text{RX},\Theta_\text{RX})=\arg\max\nolimits_{\phi_\text{RX},\theta_\text{RX}}\sum_{i=1}^L\left|h(\tau_i,\phi_\text{RX},\theta_\text{RX})\right|^2,
\end{split}
\end{equation}
\noindent where $PL_\text{O}$ is the omnidirectional PL, $PL_\text{B}$ is the PL of strongest-power direction, $\tau_i$ is the $i$th delay bin, $L$ is the number of delay bins, and $\Phi_\text{RX}$ and $\Theta_\text{RX}$ are respectively the azimuth and elevation angle of the maximum power direction.
For modeling the PL as a function of distance, the close-in (CI) model is utilized and can be expressed as:
\begin{equation}\label{CI}
\begin{split}
PL_\text{CI}(f,d)[{\rm{dB}}]=PL_\text{FSPL}(f,d_0)+10n\log_{10}(\frac{d}{d_0})+\epsilon,
\end{split}
\end{equation}
\noindent where $f$ is the center frequency of the radio wave in $\rm{GHz}$, $d_0$ is a physically-based reference distance which is set to 1 $\rm{m}$ in this paper, $n$ is the PLE, $d$ is the distance between TX and RX, and $\epsilon$ is a zero-mean Gaussian variable with a standard derivation $\sigma$ representing the shadowing, and $PL_\text{FSPL}(f,d_0)$ is the free space PL. It is written as:
\begin{equation}\label{FSPL}
\begin{split}
PL_\text{FSPL}(f,d_0)[{\rm{dB}}]=20\log_{10}(\frac{4{\pi} d_0 f}{\rm{c}}),
\end{split}
\end{equation}
\noindent where $\rm{c}$ is the speed of light. To obtain the parameters of the CI model, methods such as maximum likelihood estimation (MLE) can be used to fit the measurement results.
\subsubsection{RMS DS}\label{332}
The RMS DS is the second central moment of the PDP and is calculated as:
\begin{equation}\label{DS}
\begin{split}
\tau_\text{rms}=\sqrt{\frac{\sum\limits_{i=1}^L(\tau_{i}-\tau_\text{mean})^2\left|h_\text{omni}(\tau_{i})\right|^2}{\sum\limits_{i=1}^L\left|h_\text{omni}(\tau_{i})\right|^2}},
\end{split}
\end{equation}
with
\begin{equation}\label{mDS}
\begin{split}
\tau_\text{mean}=\frac{\sum\limits_{i=1}^L\tau_{i}\left|h_\text{omni}(\tau_{i})\right|^2}{\sum\limits_{i=1}^L\left|h_\text{omni}(\tau_{i})\right|^2}.
\end{split}
\end{equation}
\subsubsection{Angular spread}\label{333}
The measurement campaign obtains the multi-paths from 360$\degree$ in the horizontal direction with a step of the HPBW. To explore the dispersion of power over different angular directions, the angular spread should be computed. The first step is synthesizing all elevational PDPs at the same azimuth direction and getting directional angular power (DAP):
\begin{equation}\label{DPDP}
\begin{split}
DAP(\phi_\text{RX})=\max\limits_{\phi_\text{TX},\theta_\text{RX}}\sum\nolimits_{\tau}\left|h(\tau,\phi_\text{TX},\phi_\text{RX},\theta_\text{RX})\right|^2.
\end{split}
\end{equation}
Then, ASA can be computed by DAP as:
\begin{equation}\label{ASA}
\begin{split}
ASA=\sqrt{\frac{\sum\nolimits_{\phi_\text{RX}}\left|{\rm{e}}^{{\rm{j}}\phi_\text{RX}}-ASA_\text{mean}\right|^2DAP(\phi_\text{RX})}{\sum\nolimits_{\phi_\text{RX}}DAP(\phi_\text{RX})}},
\end{split}
\end{equation}
\noindent where j is the imaginary unit, and the average ASA $ASA_\text{mean}$ is written as:
\begin{equation}\label{mASA}
\begin{split}
ASA_\text{mean}=\frac{\sum\nolimits_{\phi_\text{RX}}{\rm{e}}^{{\rm{j}}\phi_\text{RX}}DAP(\phi_\text{RX})}{\sum\nolimits_{\phi_\text{RX}}DAP(\phi_\text{RX})}.
\end{split}
\end{equation}

The next section will analyze and model the PL and spread parameters based on the channel characteristics computation introduced in Section \ref{33}. Additionally, the cluster parameters, cross-correlations, and correlation distance of the spread parameters will be presented.
\section{Measurement result}\label{4}
In this section, channel characteristics are extracted, and analyzed. They are then investigated by citing the 3GPP statistical models of channel statistical characteristics \cite{r10}. It is worth mentioning that the 3GPP statistical model of channel characteristics parameters is available in the frequency bands below 100 $\rm{GHz}$. Thus, we bring the THz frequency points in the 3GPP parameters model to provide a reference for measurement results, aiming only to reveal the differences in mm-wave (below 100 $\rm{GHz}$) and THz channel propagation characteristics. Finally, the corresponding parameters for meeting the requirements of the THz channel modeling at the indoor office and the UMi scenario will be summarized.  
\subsection{PDPs}\label{41}
\begin{figure}[!t]
\centering
\subfigure[$\rm{RX10_{IL}}$ with $d$=10.1 $\rm{m}$ ]{\includegraphics[width=4.35cm]{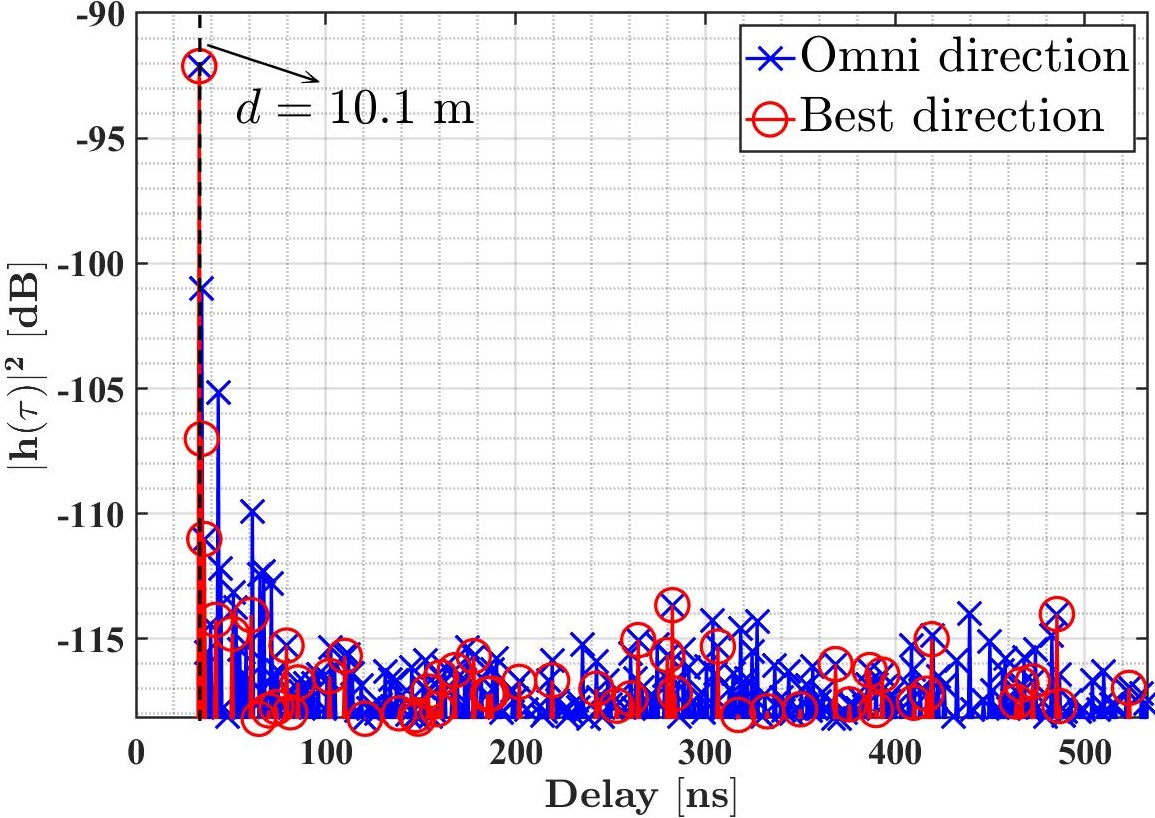}}
\subfigure[$\rm{RX4_{IN}}$ with $d$=10.9 $\rm{m}$]{\includegraphics[width=4.35cm]{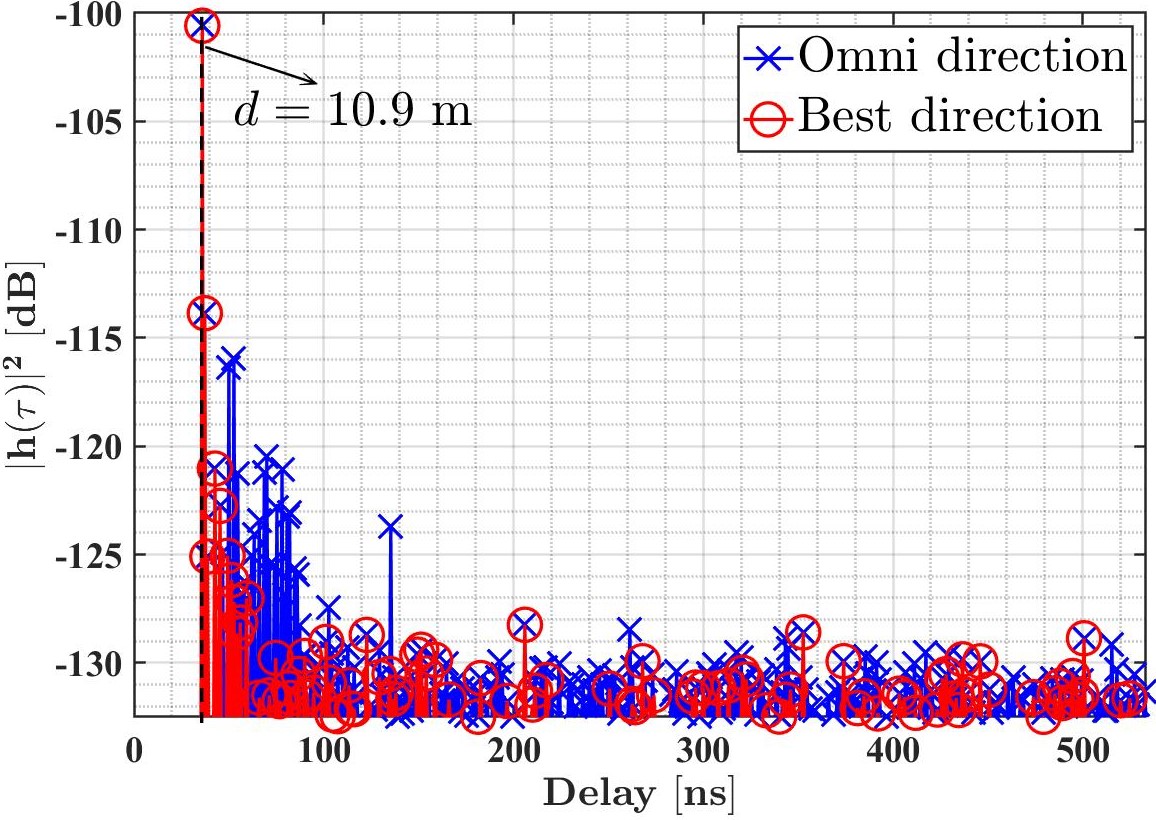}}\\
\subfigure[$\rm{RX8_{UL}}$ with $d$=44.7 $\rm{m}$]{\includegraphics[width=4.35cm]{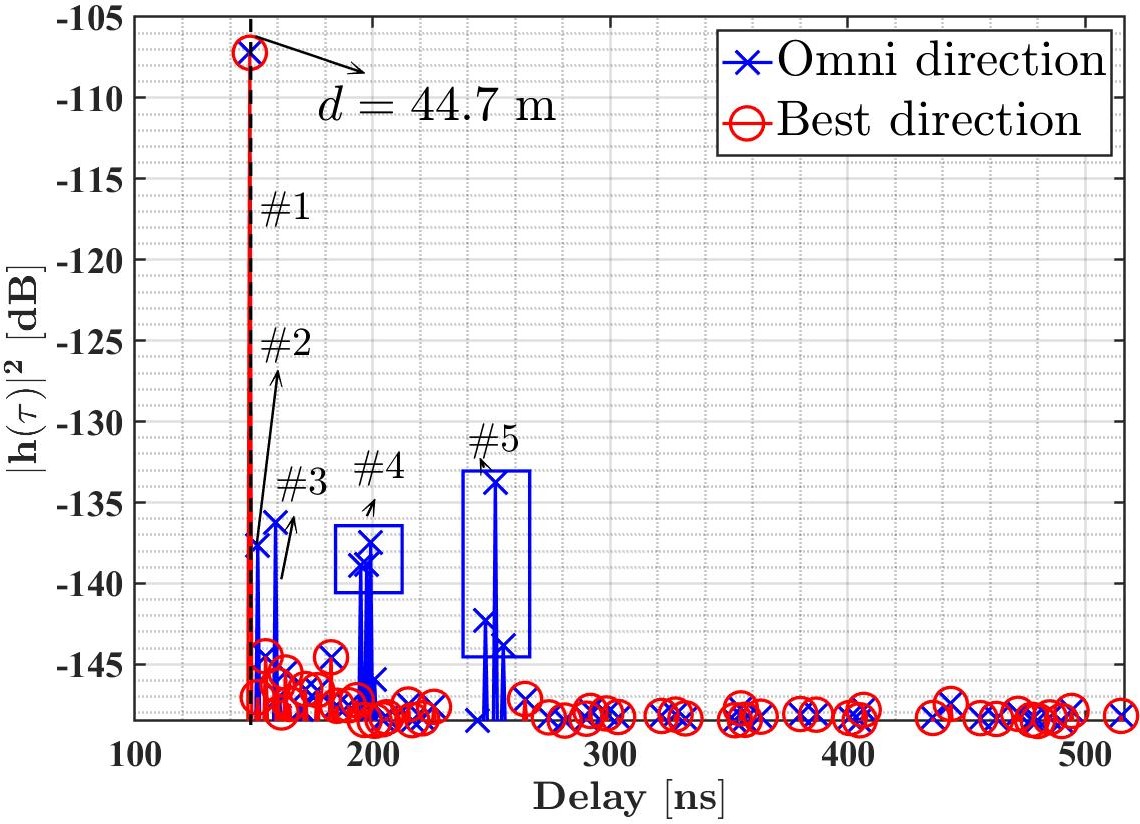}}
\subfigure[$\rm{RX8_{UN}}$ with $d$=76.4 $\rm{m}$]{\includegraphics[width=4.35cm]{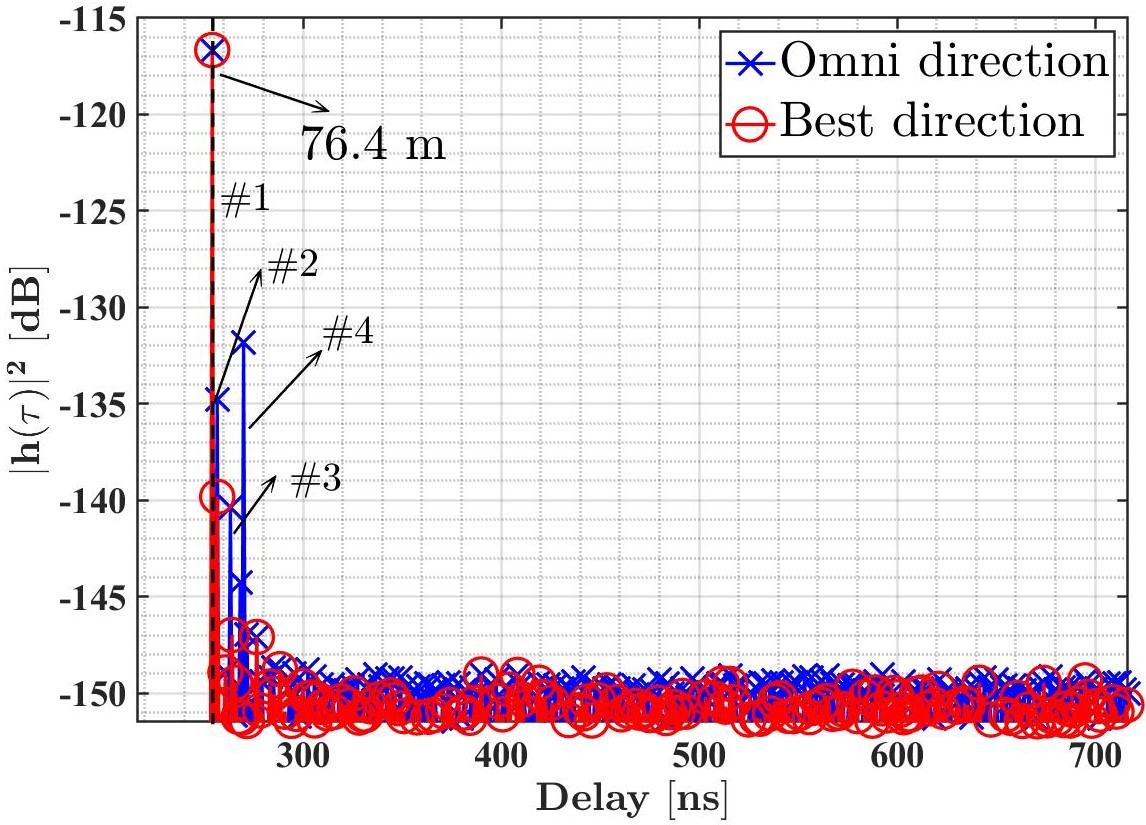}}
\caption{PDP for two positions ($\rm{RX10_{IL}}$ and $\rm{RX4_{IN}}$) in the indoor office and two positions ($\rm{RX8_{UL}}$ and $\rm{RX8_{UN}}$) in the UMi at LoS and NLoS cases, respectively}\label{Fig 7}
\end{figure}

Fig. \ref{Fig 7}a-\ref{Fig 7}b show some positions of PDPs in the indoor office, with the biggest MPC aligned at the TX-RX distance. The LoS measurement is taken at a distance of $10.1$ $\rm{m}$. Fig. \ref{Fig 7}a presents the omnidirectional and best-directional PDPs in the LoS case. The LoS path is observed in both the omnidirectional and best-directional PDPs. The second strongest MPC is $9$ $\rm{dB}$ lower than the LoS path. Besides, there are subsequent MPCs following the LoS MPC, which are not observed in the best-directional PDP. For the NLoS case, the position with a distance of $10.9$ $\rm{m}$ is shown in Fig. \ref{Fig 7}b. Apart from the biggest first reflection MPC, multiple MPCs carry power levels up to 24 $\rm{dB}$ lower than the LoS path and can only be seen in omnidirectional PDP. Moreover, the NLoS case exhibits a richer multipath environment compared to the LoS case.
\begin{figure}[!t]
\centering
\subfigure[LoS]{\includegraphics[width=4.35cm, trim=10cm 7cm 4cm 20cm, clip]{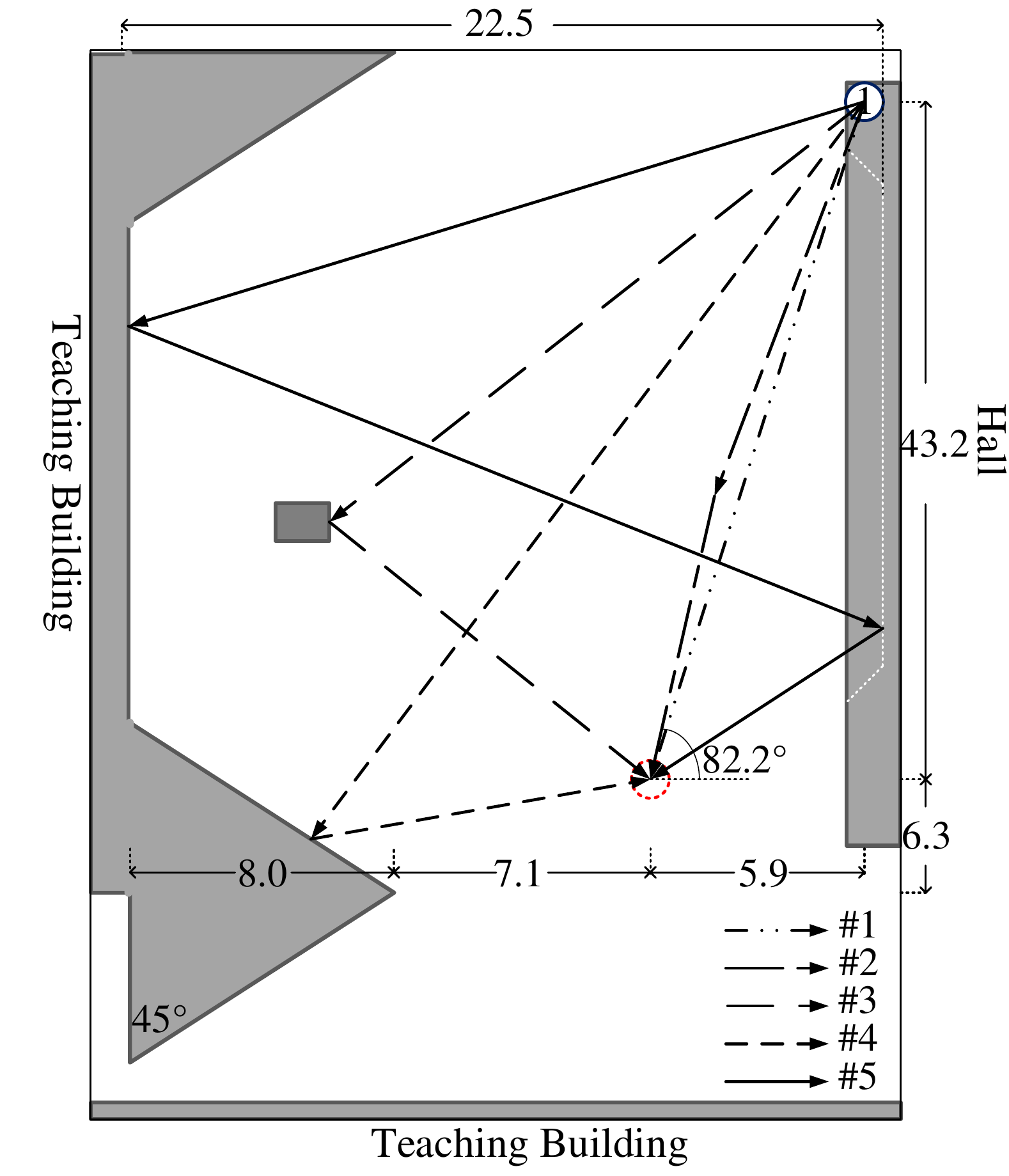}}
\subfigure[NLoS]{\includegraphics[width=4.35cm, trim=10cm 3.8cm 4cm 23.5cm, clip]{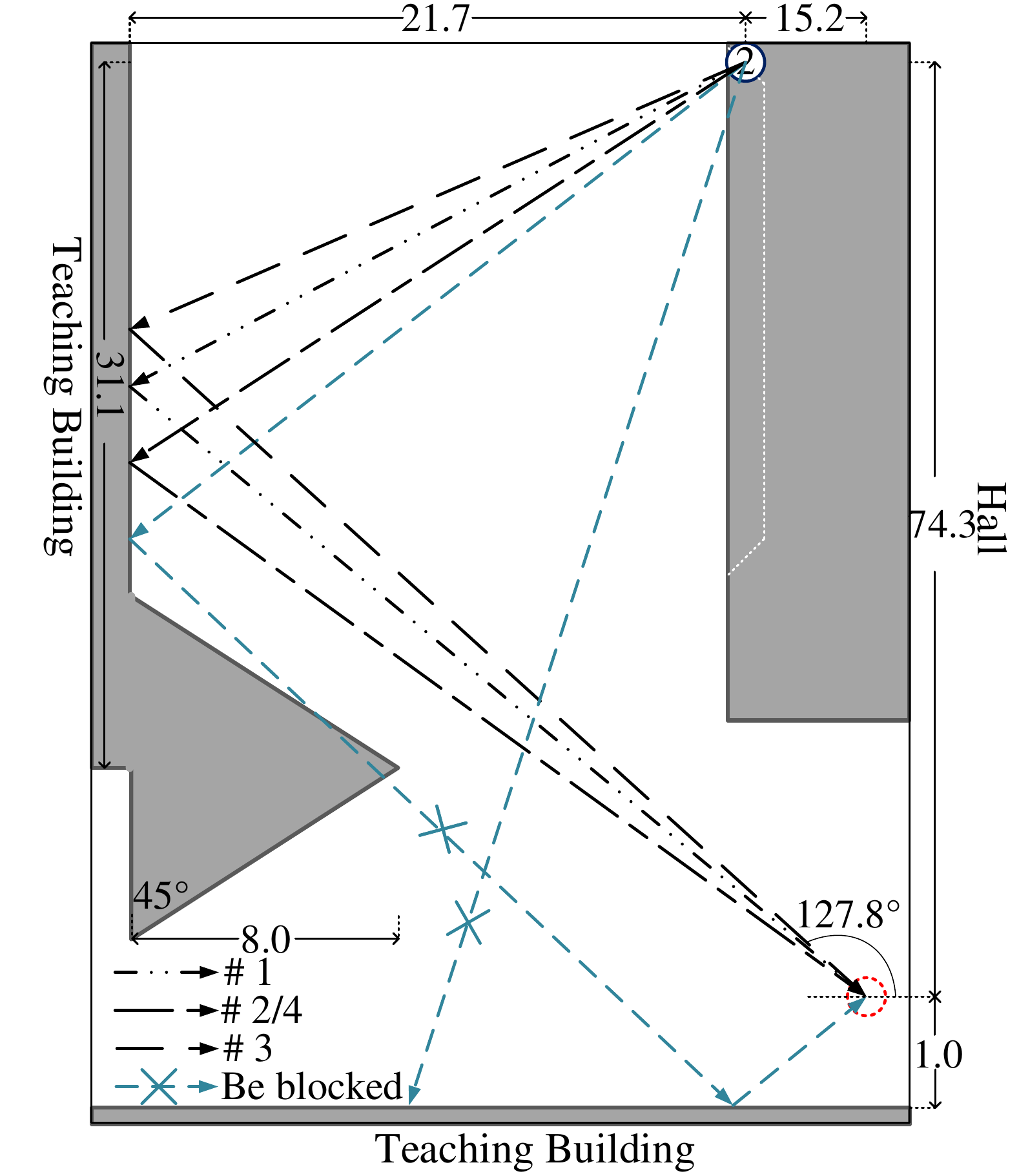}}
\caption{The propagation path analysis in the $\rm{RX8_{UL}}$ and $\rm{RX8_{UN}}$ (the unit of distance is $\rm{m}$).}\label{Fig 51}
\end{figure}

Fig. \ref{Fig 7}c-\ref{Fig 7}d presents the PDP at $\rm{RX8_{UL}}$ and $\rm{RX8_{UN}}$ in the UMi. Similarly, the LoS MPC in the LoS case or the biggest first reflection MPC in the NLoS case exists at both omnidirectional and best-directional PDPs. In the case of $\rm{RX8_{UL}}$, as shown in Fig. \ref{Fig 7}c, the omnidirectional PDP reveals multiple MPCs occurring after 100 $\rm{ns}$, with their powers being at least 27 $\rm{dB}$ lower than the LoS MPC. Compared to the indoor office, the expansive UMi scenario provides a larger channel space, leading to the generation of more MPCs with farther propagation distances in the LoS case. For the $\rm{RX8_{UN}}$ PDP presented in Fig. \ref{Fig 7}d, the number of MPCs is fewer compared to the NLoS case in the indoor office and even lower than the MPCs observed at $\rm{RX8_{UL}}$. That is because the location of the RX and the TX in the UMi scenario in the NLoS case makes it more challenging for RX to receive THz waves with reflection as the main way of propagation. 
\begin{table}[t]
\begin{center}
\setlength{\tabcolsep}{3pt}  
\footnotesize
\renewcommand\arraystretch{1.3}  
\caption{Cluster information of $\rm{RX8_{UL}}$ and $\rm{RX8_{UN}}$.}\label{Table 4}
\vglue8pt
\begin{tabular}{ccccccc}  
 \hline
   \multirow{2}{*}{\makecell{Cluster \\ number}}  & \multicolumn{3}{c}{\textbf{$\rm{RX8_{UL}}$}} & \multicolumn{3}{c}{\textbf{$\rm{RX8_{UN}}$}}\\
   \cline{2-7}
   &{AoA ($^\circ$)} & {ZoA ($^\circ$)} &{$\Delta d$ ($\rm{m}$)} &{AoA ($^\circ$)} & {ZoA ($^\circ$)} &{$\Delta d$ ($\rm{m}$)}\\
    \hline
       {$\# 1$}  &{85.2 } &{13.2} &{0.0} &{124.2} &{8.0} &{0.0}\\
    \hline
   {$\# 2$}  &{85.2} &{4.2} &{1.0} &{134.2} &{8.0} &{0.8}\\
    \hline 
     {$\# 3$}  &{145.2} &{4.2} &{3.3} &{114.2} &{8.0} &{2.8}\\
    \hline
     {$\# 4$}  &{195.2} &{13.2} &{14.0-15.3} &{134.2} &{8.0} &{4.8}\\
    \hline
     {$\# 5$}  &{45.2} &{4.2} &{29.8-32.0} &{-} &{-} &{-}\\
    \hline
\end{tabular}
\end{center}
\end{table}

Fig. \ref{Fig 51} shows the propagation path deduced according to the AoA, ZoA, and the $\Delta d$, which is the difference in propagation distance to the first arrival path, to analyze and explore the fewer clusters or MPCs in the two PDPs in the UMi. The AoA, ZoA, and $\Delta d$ are listed in Table \ref{Table 4}. For the LoS case, cluster 1 is the LoS path, and cluster 2 is from the direction of the LoS path with the $\rm{ZoA=4.2^\circ}$ and the $\Delta d=1 \ \rm{m}$ which is reflected from the ground. It is worth mentioning that the HPBW of $9^\circ$ in elevation allows the RX with the $\rm{ZoA=4.2^\circ}$ to be capable of receiving MPCs with the $\rm{ZoA<0^\circ}$. The calculated $\rm{AoA}$ of cluster 1 and cluster 2 is $\rm{82.2^\circ}$. It is within the scope of the RX HPBW with the actual received $\rm{AoA=85.2^\circ}$ ($85.2^\circ \pm 5^\circ$). Cluster 3 is received in the direction of $\rm{60^\circ}$ away from the LoS path with the $\Delta d=3.3 \ \rm{m}$. The calculated horizontal distance from the RX to the reflector for cluster 3 is 6.2 $\rm{m}$. The reflector at this location is a row of parked cars which is at the horizontal distance of 4.5-8 $\rm{m}$. Cluster 4 is reflected by the projecting part of the teaching building with the $\rm{AoA=195.2^\circ}$. The AoA of the cluster 5 is $\rm{45.2^\circ}$. The calculated propagation distance is $67.2 \ \rm{m}$ which is short by around $7 \ \rm{m}$ compared to the actual measurement in the case of only two specular reflections occurring. It means that the concave glass windows provide longer propagation distance and more reflection ways, for example, scattering. 

For the NLoS case, the clusters 1, 2, 3, and 4 are all from the direction near the maximum reflection path and the difference in propagation distance is within 4.8 $\rm{m}$. Similar to the LoS path shown in Fig. \ref{Fig 51}, the calculated $\rm{AoA}$ of cluster 1 is $\rm{127.8^\circ}$ and is also within the scope of the RX HPBW. The uneven surface of the teaching building and the large HPBW of RX and TX lead to more accessible paths around the first reflection path. However, the location of the TX and RX, and the environment prevent the RX from receiving additional AoA of MPCs. Take two cases for example. First, the teaching building at the bottom of Fig. \ref{Fig 51}b, another reflector for the two-order path, reflects the path to the left space and Rx can not receive it. Second, the three-order path reflected by both teaching buildings is blocked by the projecting part of the left teaching building. That is because the horizontal distance (8 $\rm{m}$) from the top of the projecting part to the flat surface of the left teaching building is larger than the horizontal distance from the path to the flat surface of the left teaching building which is computed as 2.6 $\rm{m}$. Overall, RX in the NLoS case is harder to receive the MPCs with different AoA and DS compared to RX in the LoS case in this UMi scenario, and THz channel exists channel sparsity due to the single propagation mode of reflection or scattering as we derived. 

Based on the PDPs, additional investigation and modeling of other channel characteristics will be carried out for the THz GBSM in the subsequent subsections. The statistical value of the channel characteristics, calculated in a manner similar to the 3GPP standard models, are summarized in Table \ref{Table 2}.
\begin{figure}[!t]
\centering
\subfigure[LoS in the indoor office]{\includegraphics[width=6.4cm]{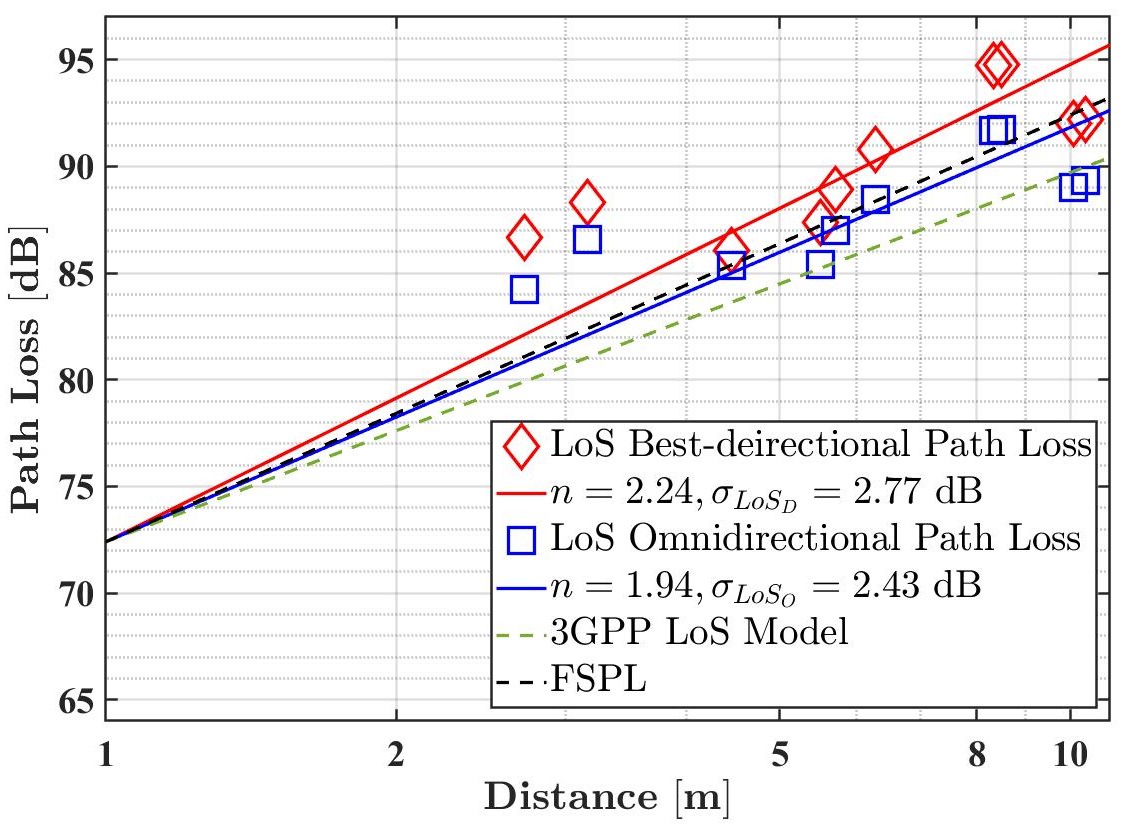}}
\subfigure[NLoS in the indoor office]{\includegraphics[width=6.4cm]{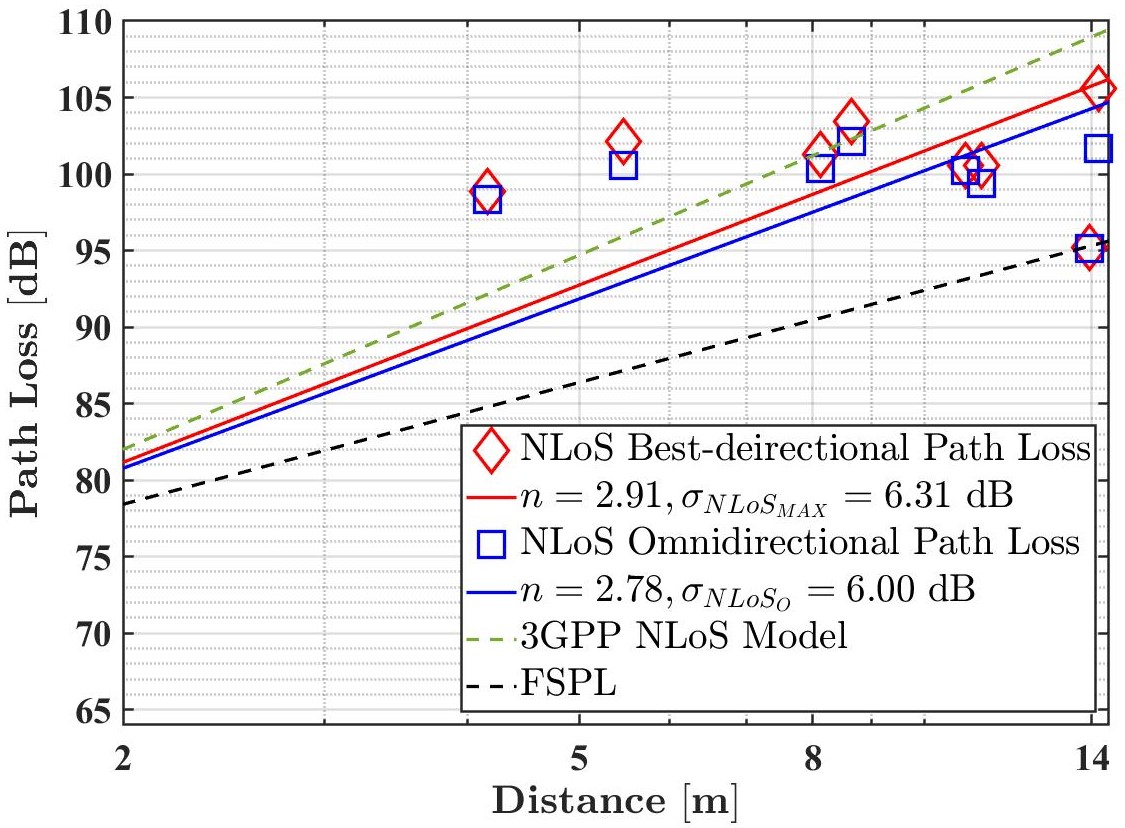}}
\caption{PL at LoS and NLoS cases for the indoor office.}\label{Fig 5}
\end{figure}
\subsection{Path loss and shadowing}\label{42}
\subsubsection{Analysis of the indoor office at 100 GHz}
Fig. \ref{Fig 5}a shows the best-directional and omnidirectional PL results in the LoS case, the FSPL model, and the 3GPP PL model with PLE of $n=1.73$ and SF standard deviation of $\sigma=3$ $\rm{dB}$ at 100 $\rm{GHz}$ \cite{r10} for the indoor office. The omnidirectional PLE is $n=1.94$ with the SF standard deviation of $\sigma=2.43$ $\rm{dB}$, which is slightly smaller than the PLE $n=2.24$ at the best direction in the LoS case. It indicates that MPCs power can be captured from directions other than boresight-aligned direction. A similar phenomenon can be observed at 142 $\rm{GHz}$ \cite{r36}. The omnidirectional PL is higher than the 3GPP PL model in the LoS case, as the MPCs in the THz channel are fewer than those in the 3GPP applicable frequency channel. 
\begin{figure}[!t]
\centering
\subfigure[LoS in the UMi]{\includegraphics[width=6.45cm]{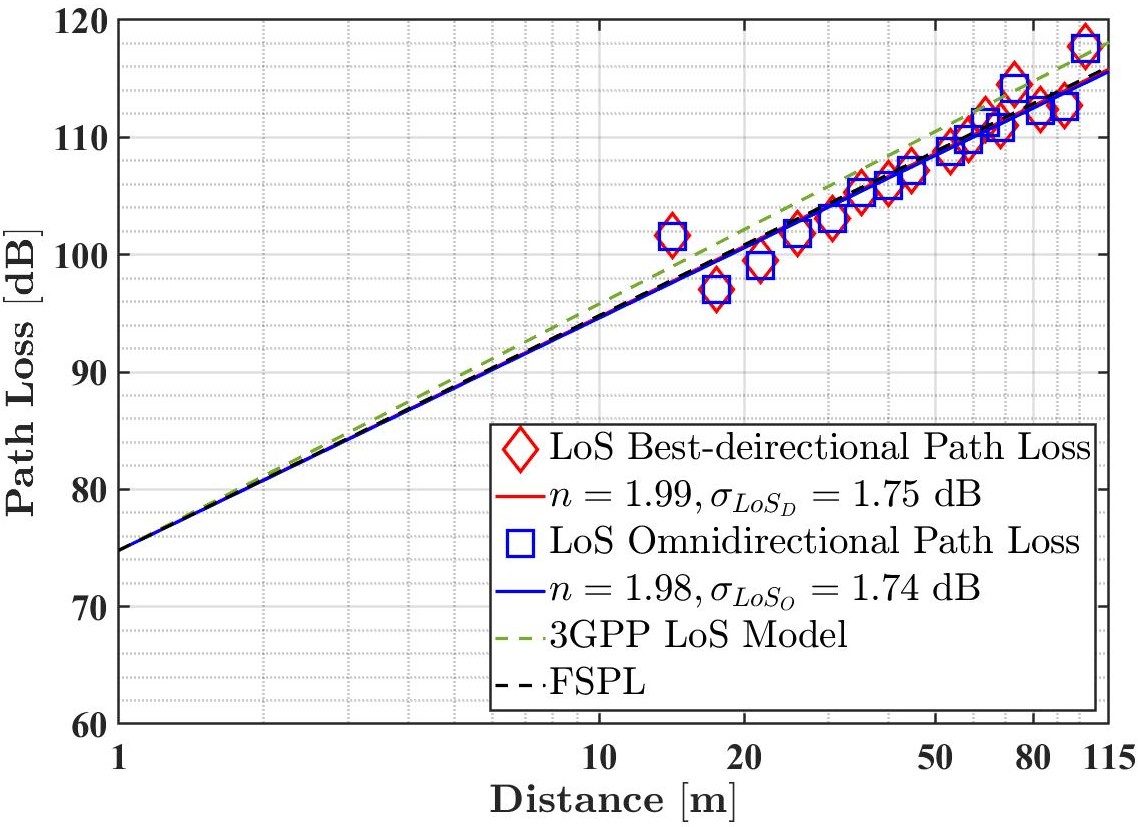}}
\subfigure[NLoS in the UMi]{\includegraphics[width=6.45cm]{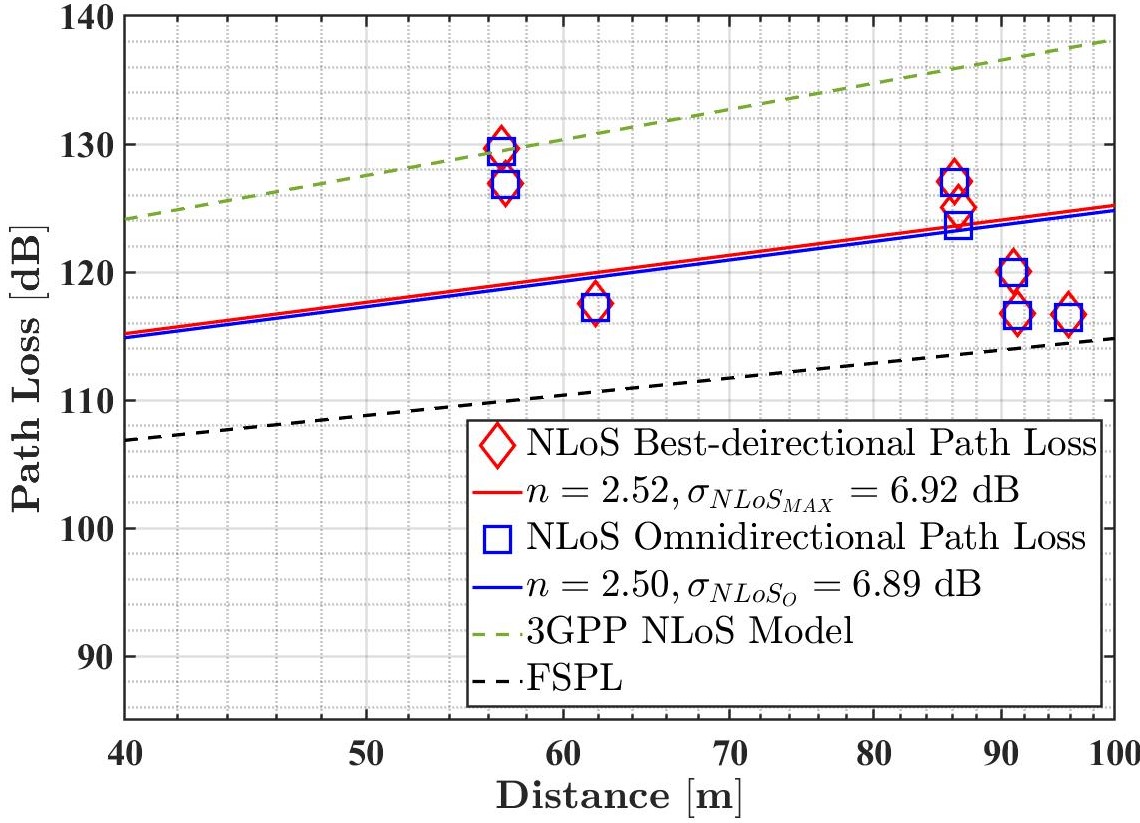}}
\caption{PL at LoS and NLoS cases for the UMi.}\label{Fig 5.5}
\end{figure}
Fig. \ref{Fig 5}b presents the best-directional and omnidirectional PL results in the NLoS case, the FSPL model, and the 3GPP PL model with $n=3.19$ and $\sigma=8.29$ $\rm{dB}$ at 100 $\rm{GHz}$ \cite{r10} for the NLoS case. The omnidirectional PLE is $n=2.78$ with the SF standard deviation of $\sigma=6.00$ $\rm{dB}$, which is smaller than but close to the PLE of $n=2.91$ in the best direction. This finding indicates that the best-directional power typically dominates among all MPCs. For specific points of the PL results, it is worth mentioning that the PL of $\rm{RX2_{IN}}$ ($d=14.0$ $\rm{m}$) in the NLoS case is close to the free space PL. This phenomenon can be attributed to the significant incident angle of the best-directional rays reflecting from TX to RX, resulting in minimal reflection and propagation losses, and the propagation distance of the path being close to the TX-RX distance $d$. A similar observation was made in the street canyon measurement at 145 $\rm{GHz}$ \cite{r24}. The results of $n=1.8$ and $\sigma=2.9$ $\rm{dB}$ in the LoS case, as well as $n=2.7$ and $\sigma=6.6$ $\rm{dB}$ in the NLoS case, are presented for comparison in the NYU WIRELESS research center \cite{140office}. Our results align with the parameters. Furthermore, many of the RX positions in the NYU measurement are located in the corridor, allowing for more captured reflection paths due to the waveguide effect. This ultimately leads to a smaller PLE. Besides, the omnidirectional PL is lower than the PL model of the 3GPP in the NLoS case. This is because the more MPCs in the lower frequency bands result in additional scattering, diffraction, and obstacle-blocking losses. 
\subsubsection{Analysis of the UMi scenario at 132 GHz}
Fig. \ref{Fig 5.5}a-Fig. \ref{Fig 5.5}b shows the best-directional and omnidirectional PL results at LoS and NLoS cases, and the FSPL model for the UMi scenario. For reference, the 3GPP PL model with $n=2.1$ and $\sigma=4$ $\rm{dB}$ in the LoS case is shown in Fig. \ref{Fig 5.5}a. Besides, Fig. \ref{Fig 5.5}b presents the 3GPP PL model in the NLoS case at 132 $\rm{GHz}$ \cite{r10}. The model is written as:
\begin{equation}\label{3gppplnlos}
\begin{split}
PL(d)[dB]=67.57+35.5\times\log_{10}(d).
\end{split}
\end{equation}

The observation of the small deviation value of 0.01 between the PLEs of the best-directional and omnidirectional model means that the received direction of the best direction contains the most power and a lot of energy other MPCs is lost due to the long propagation distance. Moreover, the measured PL is smaller than the reference 3GPP model, indicating that fewer MPCs with more power in the THz channels avoid more consumption in the large-distance transmission scenario. The path losses of $\rm{RX7_{UN}}$ and $\rm{RX8_{UN}}$ positions in the NLoS case close to the FSPL, which is consistent with the findings in the indoor office scenario. Besides, the PL tends to decrease with the increase of distance, because a larger incidence angle with distance causes a smaller reflection loss \cite{chang}, which is also observed in \cite{moliTHz}. This indicates that reflection, particularly the first reflection, is the primary propagation mechanism in the THz channel for both indoor and outdoor NLoS scenarios. Besides, the larger shadowing fading of 2.43 $\rm{dB}$ is observed in the indoor office in the LoS case than it is in the UMi. It indicates that the indoor office produces more enriched reflection paths than the UMi. The similar results of the PLEs in the LoS case are $n=1.94$ and $n=1.9$ measured in the NYU \cite{r36} and USC \cite{moliTHz}, respectively. However, the NLoS PLEs among the NYU, USC, and our measurement are discrepant. The NYU presents a higher PL with $n=2.87$, and the USC discusses a PL of up to 15 dB higher than the FSPL model which is closer to our works due to the similarity of the scenario with RXs distributed in ``street canyon". To sum up, the PL condition in both cases indicates fewer MPCs or smaller radiation range than those in the low-frequency bands, i.e., the PLEs in the LoS case close to the FSPL model and the PLs in the NLoS case tend less losses, meaning there is the channel sparsity in the THz bands.  
\begin{figure}[!t]
\centering
\subfigure[The indoor office]{\includegraphics[width=6.45cm]{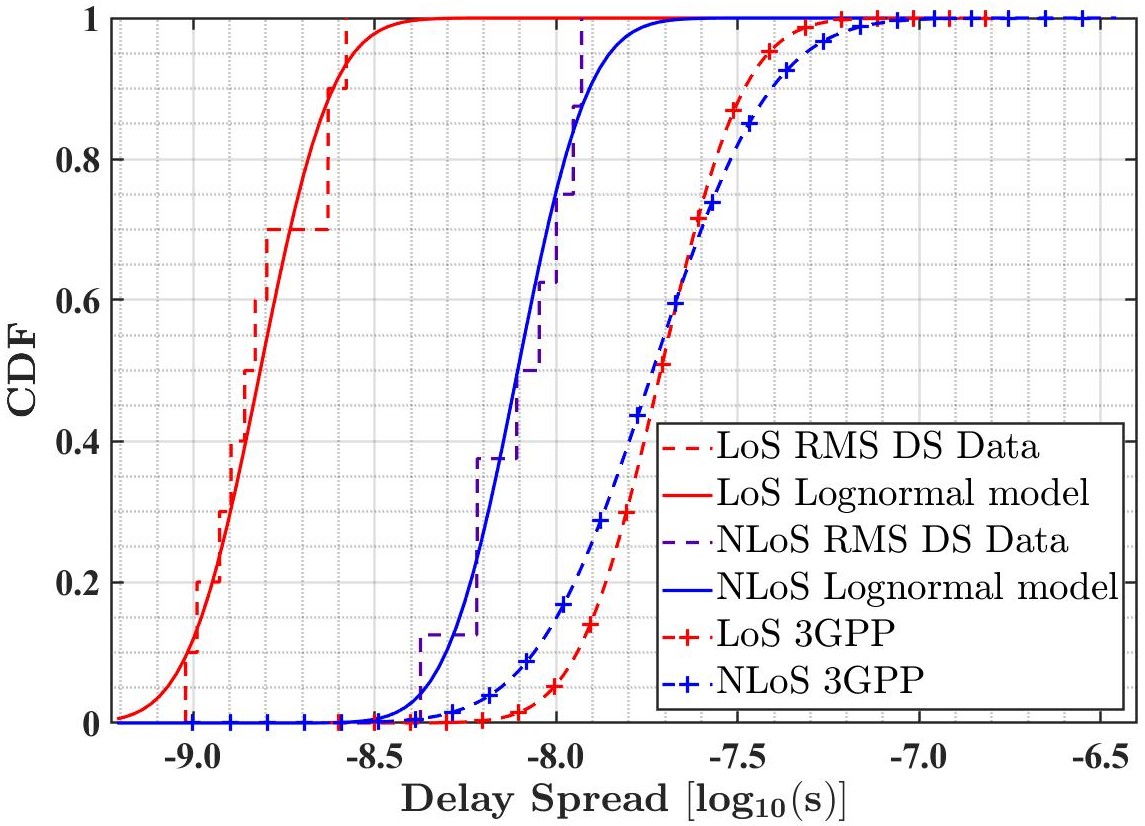}}
\subfigure[The UMi]{\includegraphics[width=6.45cm]{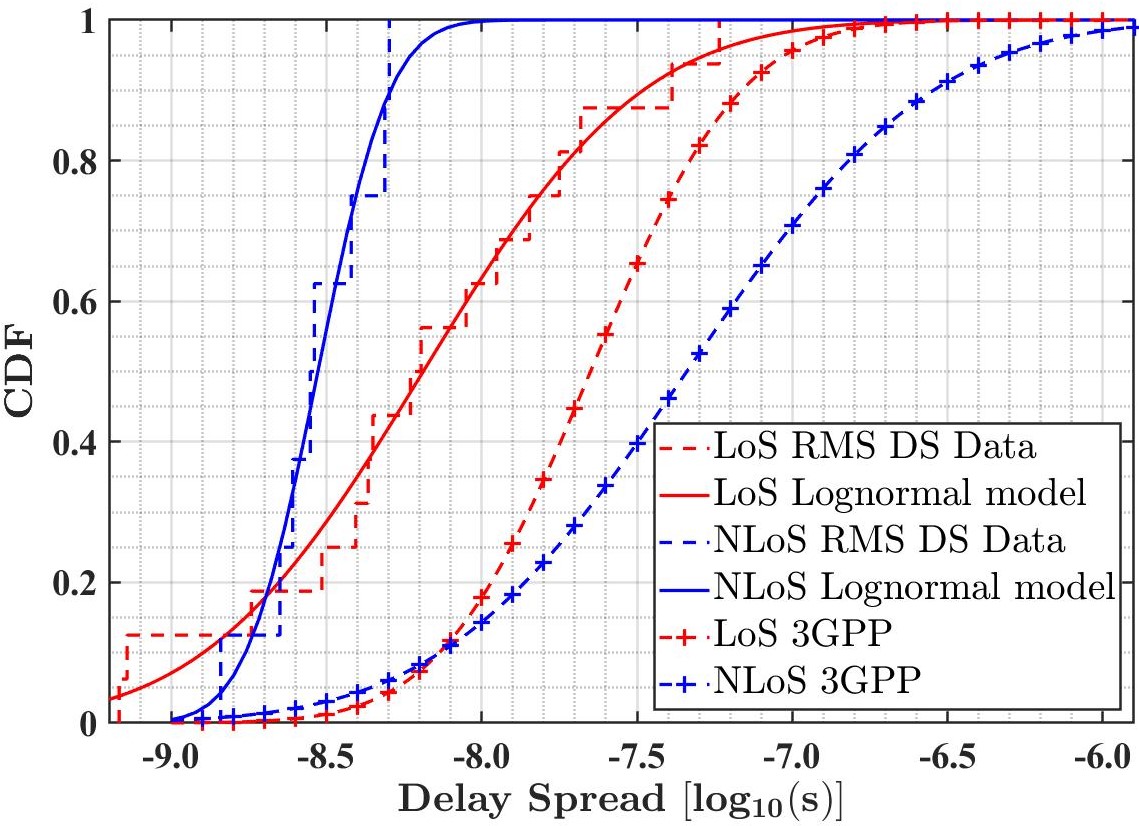}}
\caption{Normal distribution modeling of RMS DS at LoS and NLoS cases for two scenarios.}\label{Fig 8}
\end{figure}
\subsection{Delay characteristics}\label{44}
Fig. \ref{Fig 8}a presents the cumulative density function (CDF) of the measurement and 3GPP RMS DS in the indoor office. The RMS DS is plotted on a logarithmic scale, i.e., $\rm{dBs}$. The measurement RMS DS for indoor office scenario at LoS $\tau_{LoS}$ and NLoS cases $\tau_{NLoS}$ follow the lognormal distribution with $\mu=-8.82 \ \log_{10}(\rm{s})$, $\sigma=0.15 \ \log_{10}(\rm{s})$ and $\mu=-8.11 \log_{10}(\rm{s})$, $\sigma=0.15 \ \log_{10}(\rm{s})$, respectively. Notably, the RMS DS in the NLoS case is significantly larger than that in the LoS case, indicating greater spatial dispersion in NLoS cases. The NYU presents $-8.52$ and $-8.04$ $\log_{10}(\rm{s})$ as the mean value of RMS DS in the LoS case and NLoS case \cite{140office}, respectively. These values are a little bigger than the measurement results, as NYU measures a larger office. 

Fig. \ref{Fig 8}b shows the CDF of the UMi. The measurement RMS DS at LoS $\tau_{LoS}^U$ and NLoS $\tau_{NLoS}^U$ cases follow the lognormal with $\mu=-8.19 \ \log_{10}(\rm{s})$, $\sigma=0.55 \ \log_{10}(\rm{s})$ and $\mu=-8.53 \ \log_{10}(\rm{s})$, $\sigma=0.18 \ \log_{10}(\rm{s})$, respectively. The RMS DS in the NLoS case is smaller than in the LoS case. It indicates that the MPCs in the NLoS case are less than in the LoS case, which can be also observed in Fig. \ref{Fig 7}c-Fig. \ref{Fig 7}d. The presence of open space around the RX in NLoS cases results in fewer reflectors, thus reducing the number of MPCs. These results align well with those in \cite{moliTHz}, which obtains a small RMS DS in the NLoS case, and it even reaches the LoS results. 

The 3GPP RMS DS for the indoor office and UMi scenario at LoS and NLoS cases follow the lognormal distribution with $\mu=-7.71 \log_{10}(\rm{s})$, $\sigma=0.18 \ \log_{10}(\rm{s})$, $\mu=-7.73 \ \log_{10}(\rm{s})$, $\sigma=0.26 \ \log_{10}(\rm{s})$, $\mu=-7.65 \ \log_{10}(\rm{s})$, $\sigma=0.38 \ \log_{10}(\rm{s})$, and $\mu=-7.34 \ \log_{10}(\rm{s})$, $\sigma=0.62 \ \log_{10}(\rm{s})$, respectively \cite{r10}. The measured RMS DS of both scenarios is smaller than those of the 3GPP channel model which suggests the MPCs at the THz channel have shorter propagation distances. For the aspect on the delay domain, the THz channel exhibits sparser MPCs than the low-frequency channel and becomes sparser. 
\begin{figure}[!t]
\centering
\subfigure[The indoor office]{\includegraphics[width=6.45cm]{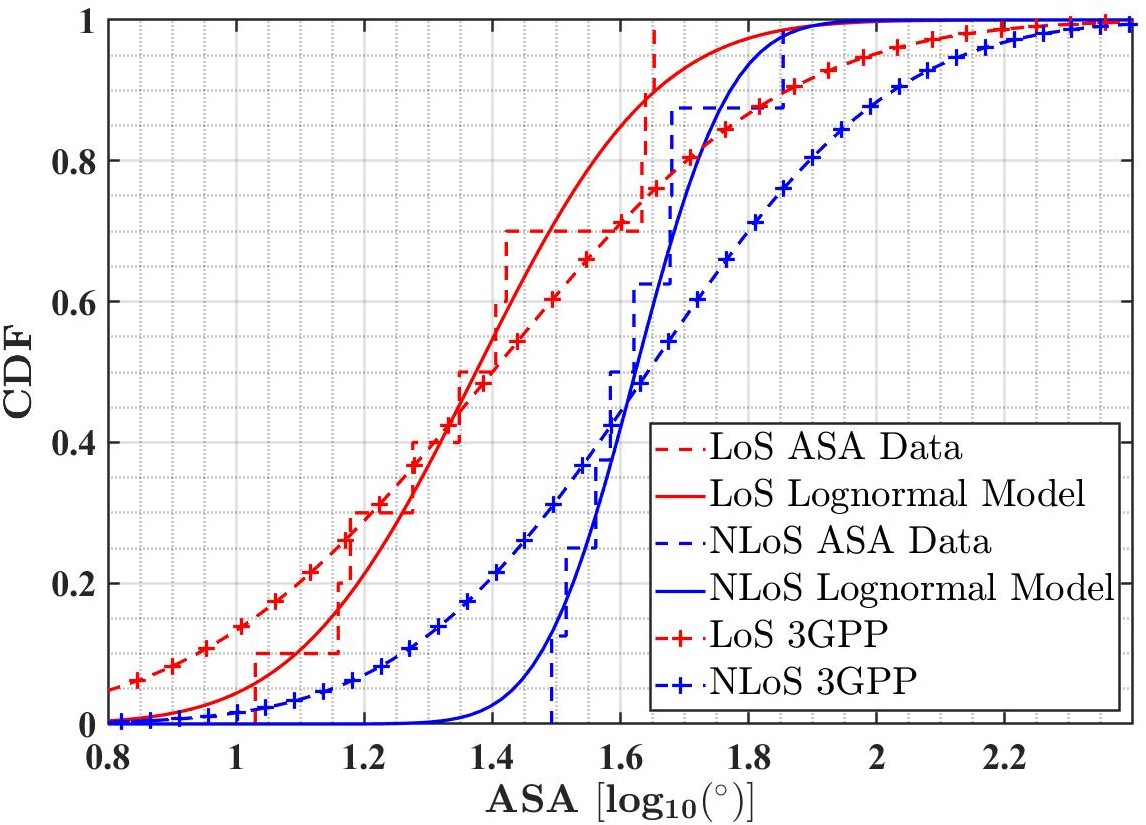}}
\subfigure[The UMi]{\includegraphics[width=6.45cm]{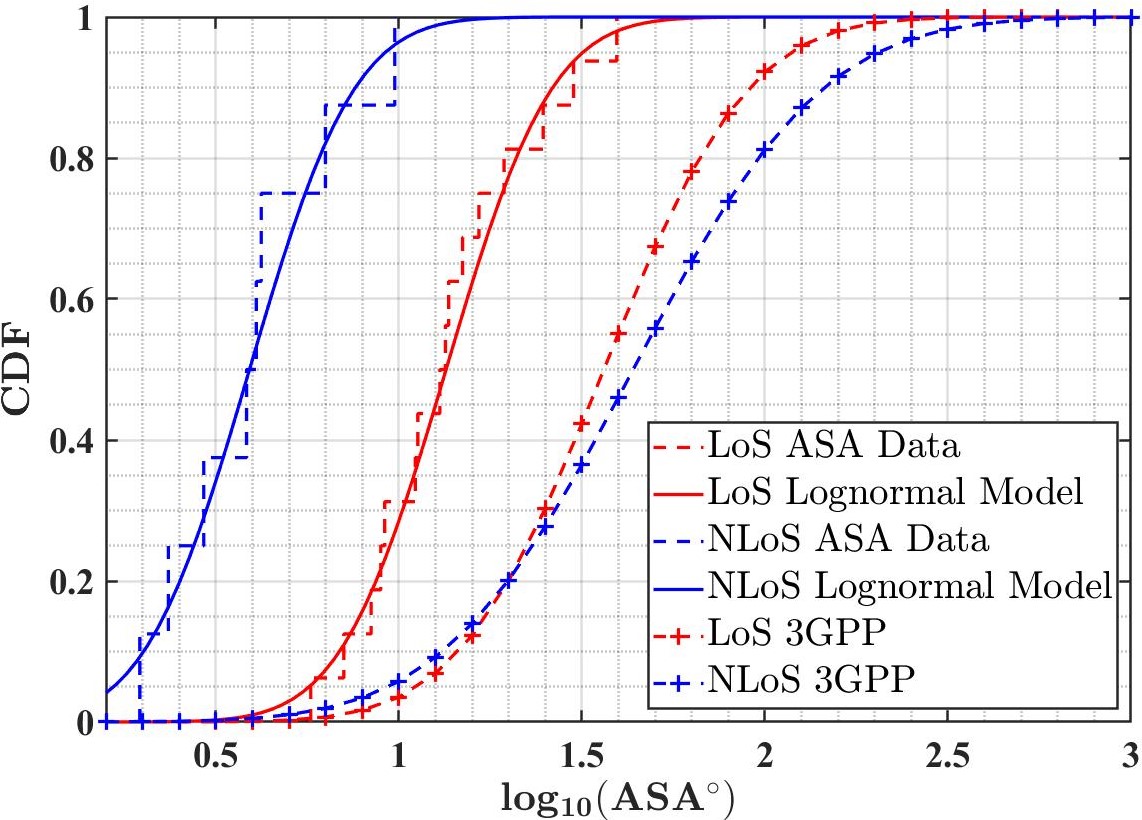}}
\caption{Normal distribution Modeling of ASA at LoS and NLoS cases for two scenarios.}\label{Fig 9}
\end{figure}
\subsection{Azimuth angular characteristics}\label{45}
Fig. \ref{Fig 9}a shows the ASA results of the measurement and the 3GPP at 100 GHz. The measured ASA at LoS and NLoS cases in the indoor office follow the lognormal distribution with $\mu=1.37 \ \log_{10}(\degree)$, $\sigma=0.21 \ \log_{10}(\degree)$ and $\mu=1.62 \ \log_{10}(\degree)$, $\sigma=0.11 \ \log_{10}(\degree)$, respectively. The NLoS channel has more MPCs from multiple angles, so the NLoS case has a larger ASA than the LoS case. In \cite{r19}, the similarity values of ASA are $1.46$ and $1.69 \ \log_{10}(\degree)$ for LoS and NLoS case, respectively. Furthermore, \cite{140office} reveals a smaller ASA, with the mean ASA in the NLoS case being 0.73 $\log_{10}(\degree)$ and smaller than that in the LoS case, attributed to the specific RX setup mentioned in subsection \ref{42}.

Fig. \ref{Fig 9}b shows the ASA results of the measurement and the 3GPP at 132 GHz. The measurement ASA at LoS and NLoS cases in the UMi follow the lognormal distribution with $\mu=1.13 \ \log_{10}(\degree)$, $\sigma=0.23 \ \log_{10}(\degree)$ and $\mu=0.59 \ \log_{10}(\degree)$, $\sigma=0.23 \ \log_{10}(\degree)$, respectively. The ASA in the NLoS case is smaller than that in the LoS case. That is because the RX in the NLoS channel only receives the MPCs around the first reflection direction. The major power of the THz channel in the UMi is received from a narrow arrival angle as shown in Fig. \ref{Fig 51}b. A lot of paths after the first reflection are reflected in the large space and can not arrive at the RX according to the geometric theory. This phenomenon has also been observed in \cite{140office} and \cite{moliTHz}. The mean ASAs in \cite{moliTHz} are even less than $0 \ \log_{10}(\degree)$ at LoS and NLoS case. 

Besides, the 3GPP ASA for the indoor office and UMi scenario in the LoS and NLoS cases follow the lognormal distribution with $\mu=1.40 \ \log_{10}(\degree)$, $\sigma=0.36 \ \log_{10}(\degree)$, $\mu=1.64 \ \log_{10}(\degree)$, $\sigma=0.30 \ \log_{10}(\degree)$, $\mu=1.56 \ \log_{10}(\degree)$, $\sigma=0.31 \ \log_{10}(\degree)$, and $\mu=1.64 \ \log_{10}(\degree)$, $\sigma=0.41 \ \log_{10}(\degree)$, respectively. The measured ASA exhibits a higher concentration around the average value than the reference distribution. This can be attributed to the fact that THz propagation primarily occurs through reflection, while the mm-wave channel, as defined by 3GPP, involves multiple propagation mechanisms such as reflection, transmission, and diffraction. Consequently, the reflection dominating propagation in the THz channel makes it hard to receive MPCs. Thus, the THz channel experiences fewer receiving angles of MPCs and a lower occurrence of ASA compared to the mm-wave channel and suffers channel sparsity for the spatial domain.
\subsection{K-factor}\label{46}
\begin{figure}[!t]
\centering
\subfigure[The indoor office]{\includegraphics[width=6.45cm]{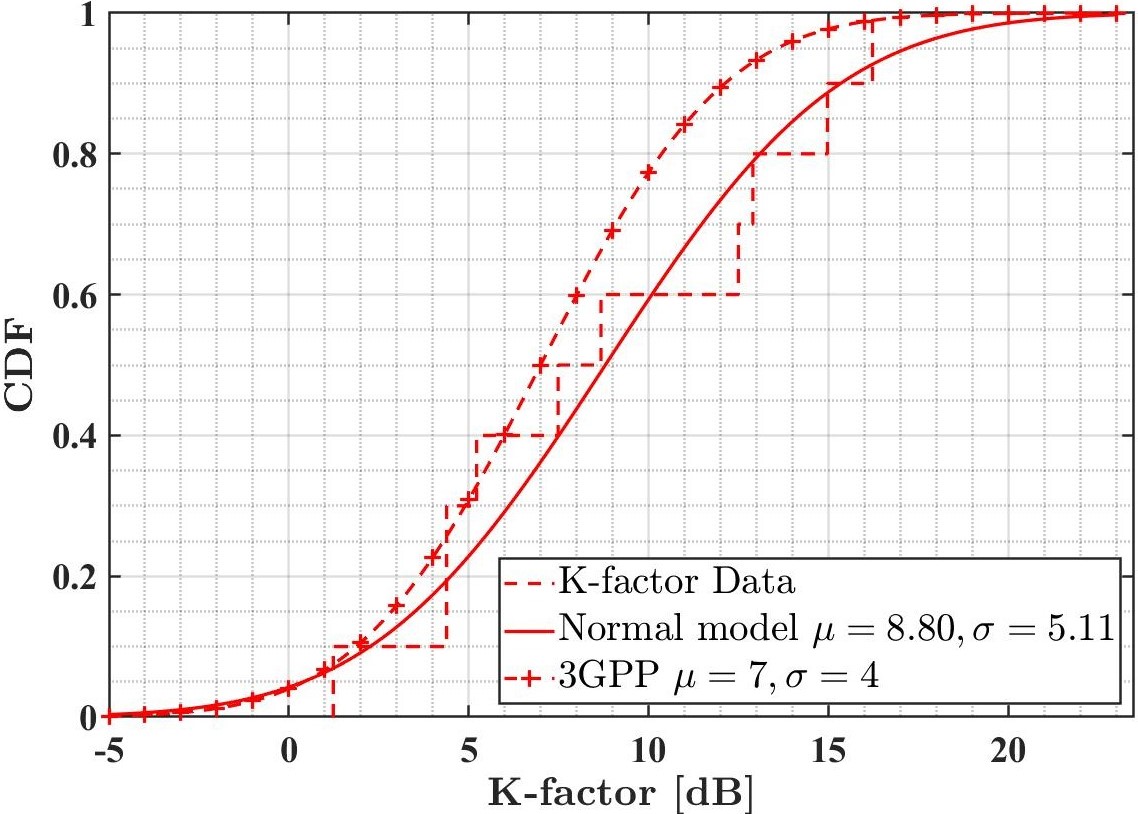}}
\subfigure[The UMi]{\includegraphics[width=6.45cm]{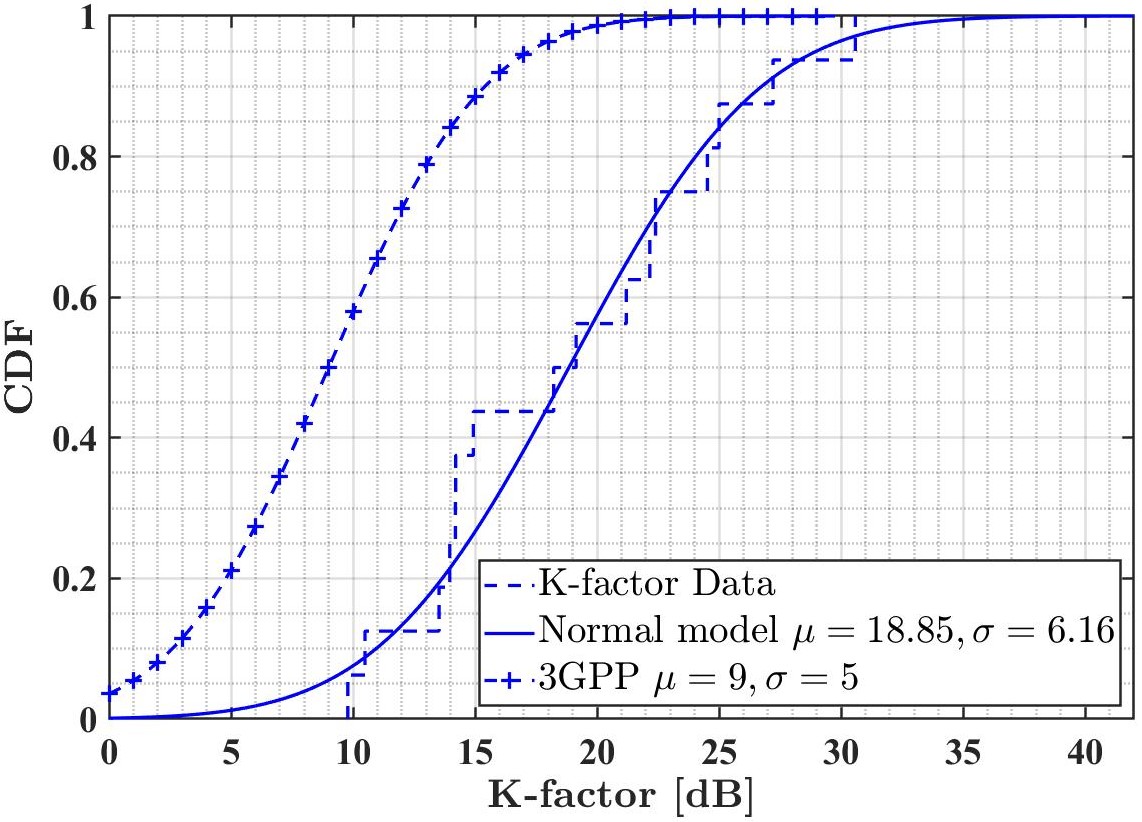}}
\caption{Modeling for K-factor for two scenarios.}\label{Fig 10}
\end{figure}

In this subsection, the K-factor of the measurement and 3GPP for two scenarios are analyzed in Fig. \ref{Fig 10}. It is important to analyze the distribution of the MPC over the delay domain \cite{kf}. The K-factor $K_\text{R}$ is defined as the ratio of the strongest MPC to other MPCs in the LoS case \cite{ttk}. Fig. \ref{Fig 10}a shows the K-factor of the measurement and 3GPP \cite{r10} in the indoor office in $\rm{dB}$, which follow the normal distribution with $\mu=8.80 \ \rm{dB}$, $\sigma=5.11  \ \rm{dB}$ and $\mu=7  \ \rm{dB}$, $\sigma=4  \ \rm{dB}$, respectively. The measured K-factor is bigger than that in the 3GPP because the THz channel has fewer MPCs, and the power of the LoS path dominates among all MPCs. 

The modeling results of the measurement in the UMi are presented in Fig. \ref{Fig 10}b. Similarly, the measured K-factor in $\rm{dB}$ in the UMi with the normal distribution of $\mu=18.85 \ \rm{dB}$, $\sigma=6.16  \ \rm{dB}$ is bigger than the default values in the 3GPP channel model with the normal distribution of $\mu=9  \ \rm{dB}$, $\sigma=5  \ \rm{dB}$. Besides, the K-factor in the UMi is larger than that in the indoor office. The reason is that the MPCs in the UMi with the long propagation distance have smaller power than those in the indoor office. Thus, the power in the UMi is more focused on the dominant path.
\subsection{Cluster characteristics}\label{47}
The cluster is the basic unit that constitutes the channel space-time propagation characteristics. We employ the K-power-means algorithm to cluster the MPCs, with cluster centroids iteratively adjusted based on the multiple component distance (MCD) \cite{r25}. The statistical results for cluster mean numbers, cluster DS, cluster ASA, and cluster K-factor are calculated and shown in Table \ref{Table 2}. 

\begin{figure}[!t]
\centering
\subfigure[The indoor office]{\includegraphics[width=6.45cm]{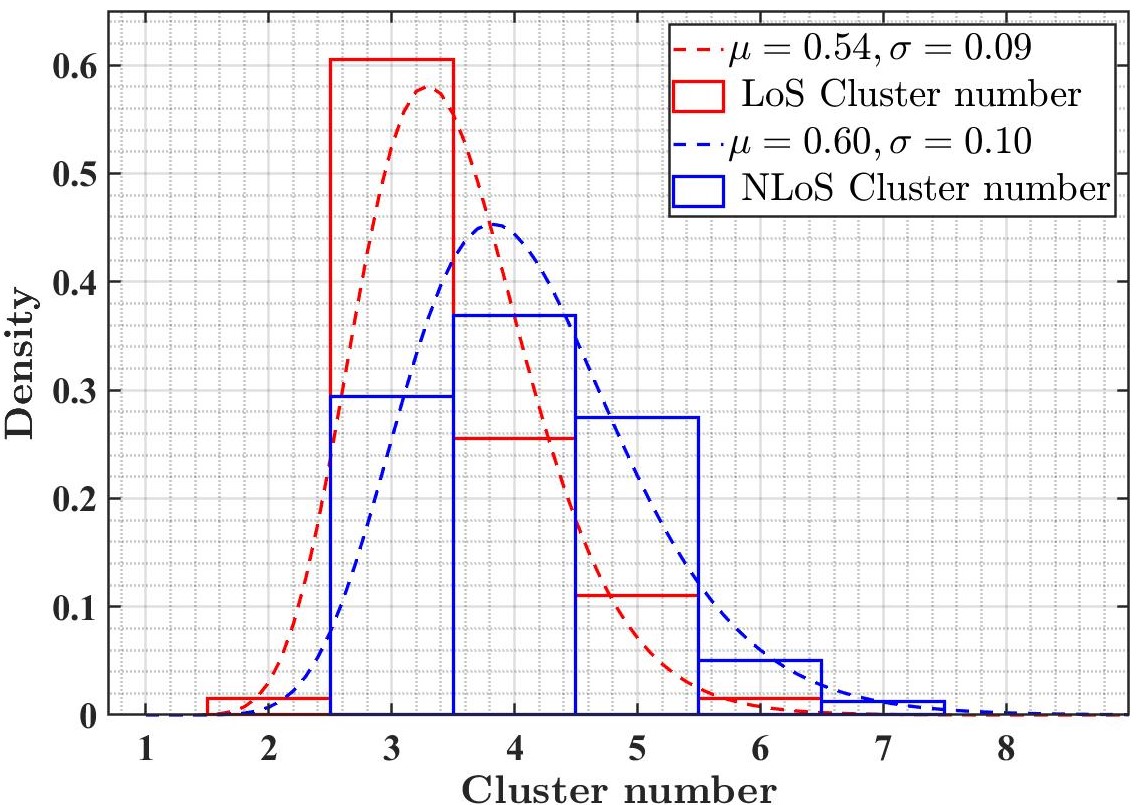}}
\subfigure[The UMi]{\includegraphics[width=6.45cm]{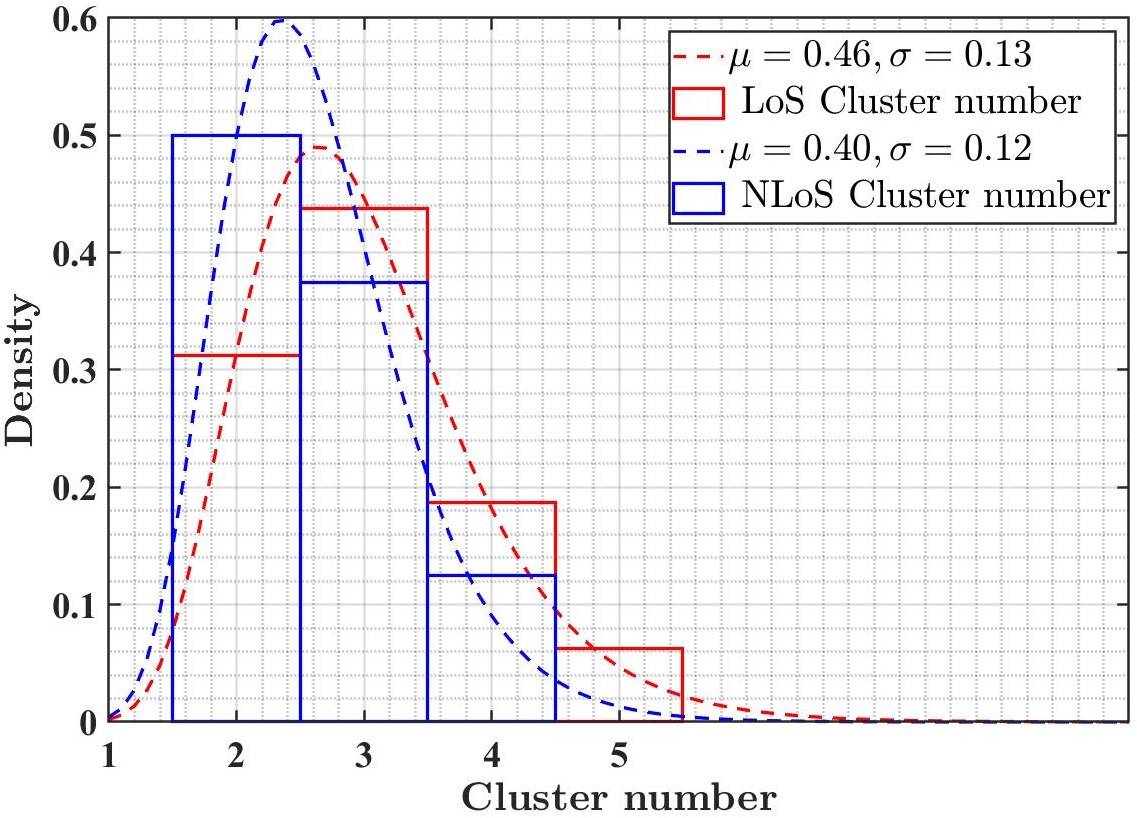}}
\caption{The PDF of cluster numbers for two scenarios.}\label{Fig 11}
\end{figure}

\begin{figure}[!t]
\centering
\subfigure[$\rm{RX10_{IL}}$]{\includegraphics[width=4.35cm]{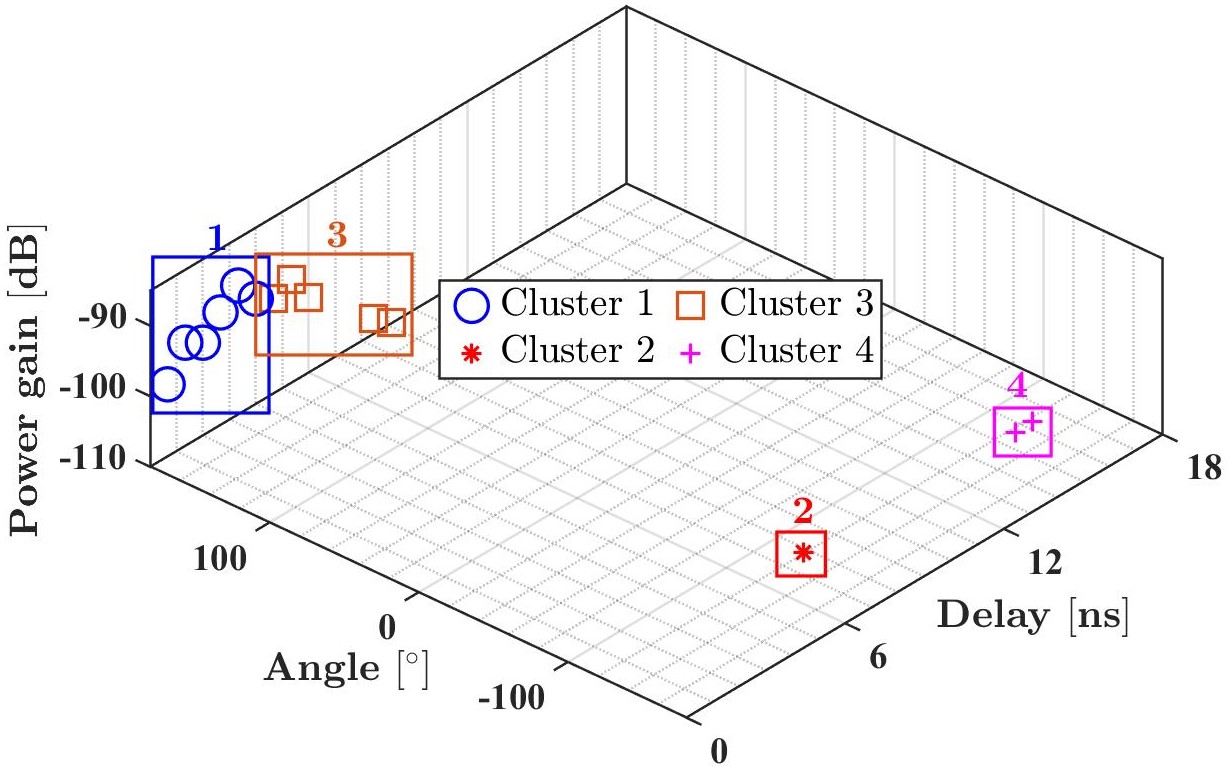}}
\subfigure[$\rm{RX6_{IN}}$]{\includegraphics[width=4.35cm]{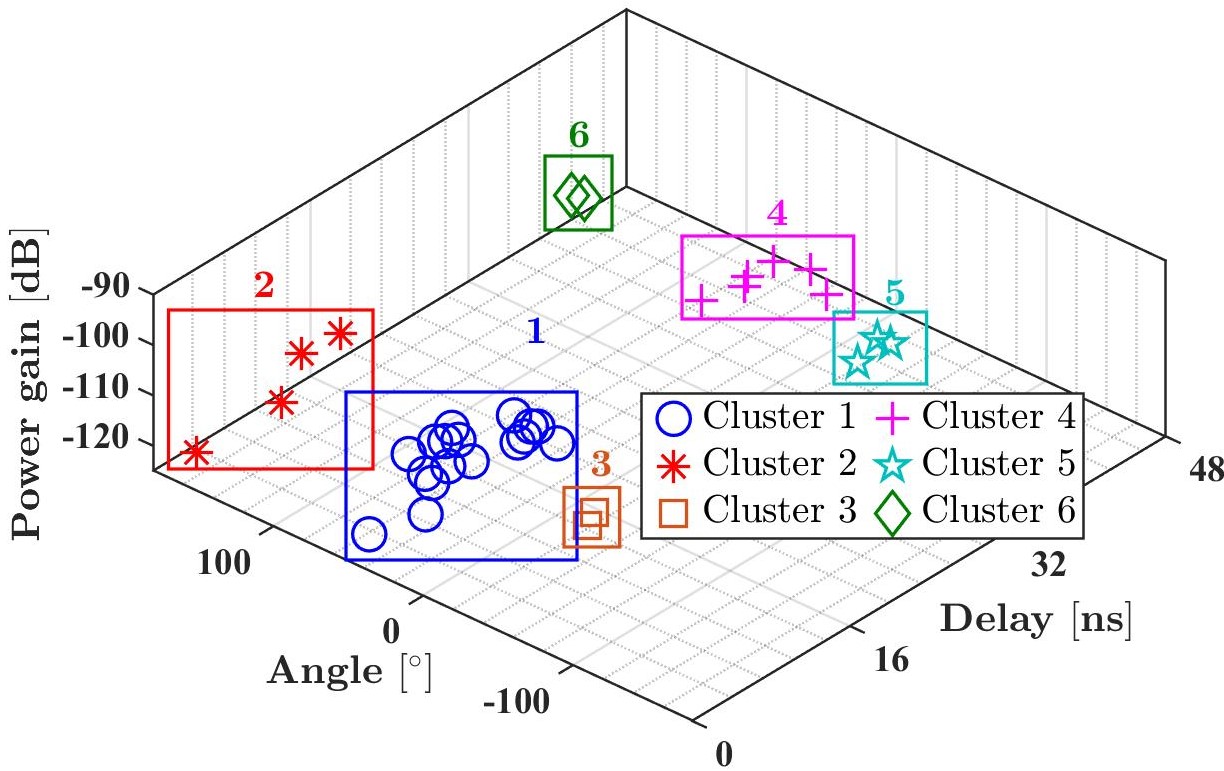}}\\
\subfigure[$\rm{RX8_{UL}}$]{\includegraphics[width=4.35cm]{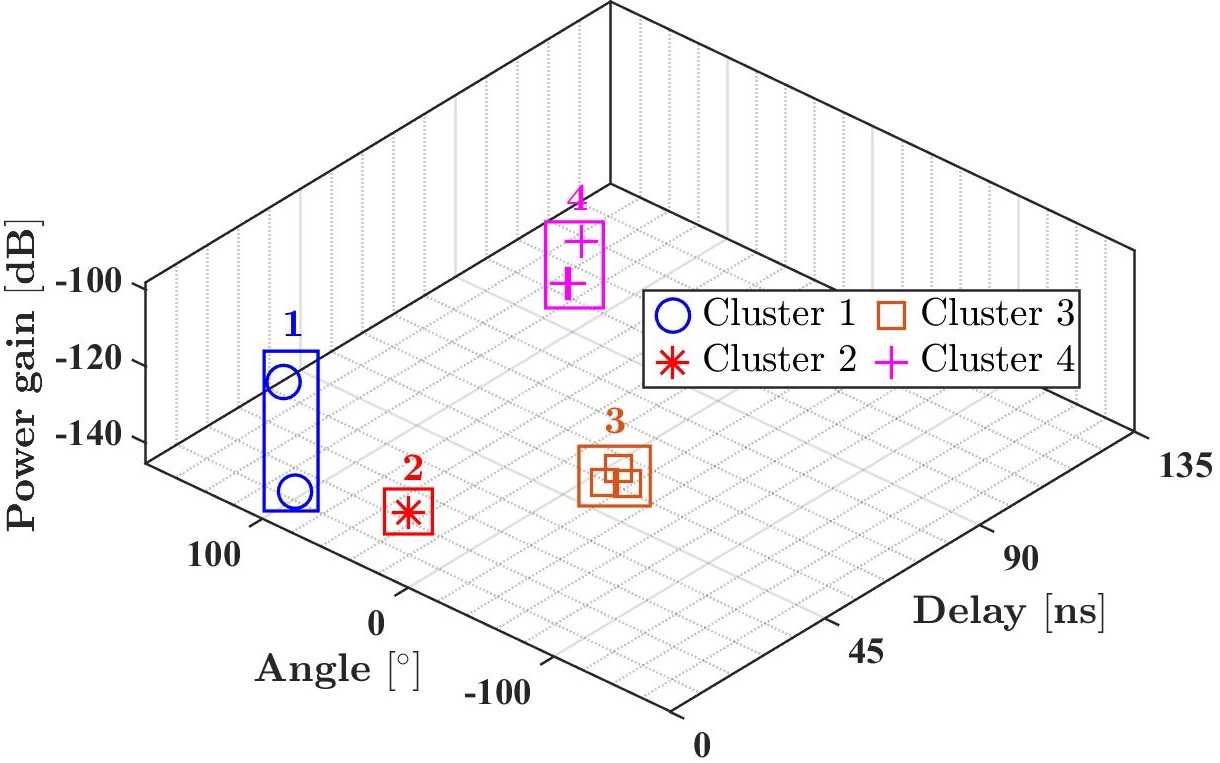}}
\subfigure[$\rm{RX1_{UN}}$]{\includegraphics[width=4.35cm]{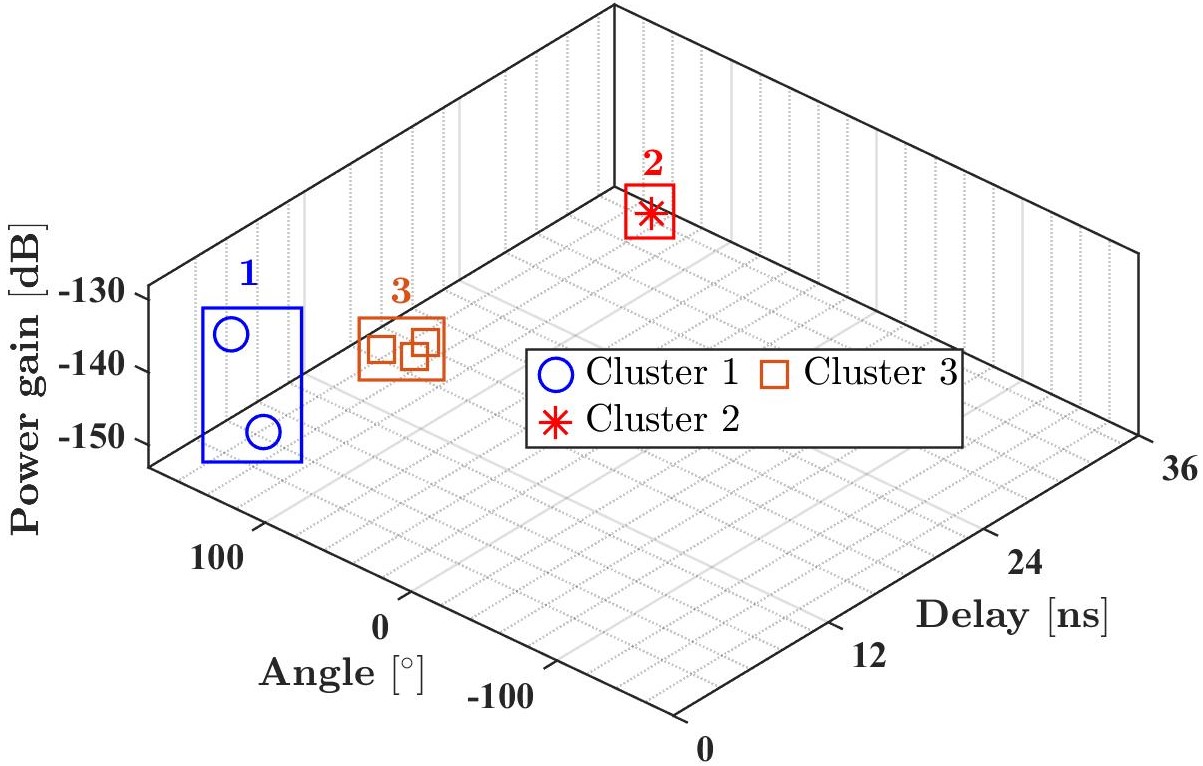}}
\caption{The clustering result at $\rm{RX10_{IL}}$ and $\rm{RX6_{IN}}$ in the indoor office and at $\rm{RX3_{UL}}$ and $\rm{RX1_{UN}}$ in the UMi.}\label{Fig 11.5}
\end{figure}

For the indoor office, the DS and ASA in the LoS case are smaller compared to the NLoS case, indicating that the MPCs within a cluster exhibit greater similarity in terms of delay and angle. The cluster numbers in the LoS case are less than those in the NLoS case. Fig. \ref{Fig 11}a shows the results of statistical fitting for the cluster numbers in the indoor office at LoS and NLoS cases. Furthermore, the cluster numbers follows lognormal distribution with $\mu=0.54$, $\sigma=0.09$ and $\mu=0.60$, $\sigma=0.10$ at LoS and NLoS cases, respectively. The cluster numbers of the same scenario are similar and present a smaller-variance lognormal distribution which is different from the traditional Poisson model at mm-wave \cite{r26}. The reason is that reflection dominates the THz propagation mechanism and the small indoor office scenario provides a simple reflector layout. Fig. \ref{Fig 11.5}a-\ref{Fig 11.5}b illustrate the clustering results at $\rm{RX10_{IL}}$ in the LoS case and $\rm{RX6_{IN}}$ in the NLoS case, respectively. The first arriving path is set at 0 $\rm{ns}$. The clusters in the LoS case are almost from two angles. These two angles differ by 60$^\circ$ while the clusters in the NLoS case are from about three directions involving 180$^\circ$. The NLoS case has more MPCs and the signal can cover more directions. Besides, it is found that more than one cluster is in similar angular intervals. That is because the wide beamwidth of TX creates large-scale MPCs with different propagation distances to be reflected. 

For the UMi, the DS, ASA, and cluster numbers in the LoS case are bigger, which indicates that the street canyon provides more MPCs with different long propagation distances in the LoS case. Besides, it is difficult to generate paths with abundant AoAs and distances in the NLoS case in the UMi. It is found that the DS is bigger than that in the indoor office in the LoS case. That is because the propagation distance of MPCs is longer in the UMi. Furthermore, the cluster numbers follows lognormal distribution with $\mu=0.46$, $\sigma=0.13$ and $\mu=0.40$, $\sigma=0.12$ at LoS and NLoS cases, respectively. The clustering results at $\rm{RX8_{UL}}$ in the LoS case and $\rm{RX1_{UN}}$ in the NLoS case are shown in Fig. \ref{Fig 11.5}c-\ref{Fig 11.5}d. The first arriving path is set at 0 $\rm{ns}$. The reflection clusters in the LoS case contain three clusters (clusters 2, 3, and 4) at the AoA of 60$^\circ$, 110$^\circ$ and 40$^\circ$ away from the LoS clusters, respectively, while almost all clusters are at AoA around the first-reflection directions in the NLoS case. The farthest spreading clusters are approximately 100 $\rm{ns}$ (equivalent to 30 $\rm{m}$) in the LoS case and 30 $\rm{ns}$ (equivalent to 9 $\rm{m}$) in the NLoS case away from the nearest spreading cluster. The clusters in the NLoS case are all reflected by the uneven surface of the left teaching building.   
\begin{table*}[t]
\begin{center}
\setlength{\tabcolsep}{5.5pt}  
\footnotesize
\renewcommand\arraystretch{1.3}  
\caption{THE 3GPP-LIKE CHANNEL PARAMETERS OF THE MEASUREMENT.}\label{Table 2}
\vglue8pt
\begin{tabular}{cccccc}  
 \hline
  \multicolumn{2}{c}{\textbf{Scenarios}}   &{\textbf{Office LoS}} &{\textbf{Office NLoS}} &{\textbf{UMi LoS}} &{\textbf{UMi NLoS}}\\     
 \hline
 {PL (CI model)}  &{PLE}   &{1.94}    &{2.78} &{1.98}    &{2.50}\\
  \hline
 \multirow{2}{*}{\makecell{DS \\ $\lg{{\rm{DS}}}$=$\log_{10}{{{\rm{(DS/1 s)}}}}$}} &{$\mu_{\lg{\rm{DS}}}$}   &{-8.82}    &{-8.11} &{-8.19}    &{-8.53}\\
 \cline{2-6}
 &{$\sigma_{\lg{\rm{DS}}}$}   &{0.15}    &{0.15} &{0.55}    &{0.18}\\
 \hline
 \multirow{2}{*}{\makecell{ASA \\ $\lg{{\rm{ASA}}}$=$\log_{10}{{{\rm{(ASA/1 \degree)}}}}$}} &{$\mu_{\lg{\rm{ASA}}}$}   &{1.37}    &{1.62} &{1.13}    &{0.59}\\
 \cline{2-6}
 &{$\sigma_{\lg{\rm{ASA}}}$}   &{0.21}    &{0.11} &{0.23}    &{0.23}\\
 \hline
  {SF [$\rm{dB}$]}  &{$\sigma_{SF}$}   &{2.43}    &{6.00} &{1.74}    &{6.89}\\
  \hline
 \multirow{2}{*}{K [$\rm{dB}$]} &{$\mu_{K}$}   &{8.80}    &{-} &{18.85}    &{-}\\
 \cline{2-6}
 &{$\sigma_{K}$}   &{5.11}    &{-} &{6.16}    &{-}\\
 \hline
 \multirow{6}{*}{Cross-Correlations} &{ASA vs DS} &{0.10} &{0.33} &{0.45} &{-0.42}\\
  \cline{2-6}
  &{ASA vs SF} &{0.38} &{-0.57} &{-0.30} &{0.10}\\
  \cline{2-6}
  &{DS vs SF} &{0.47} &{-0.49} &{-0.10} &{0.56}\\
  \cline{2-6}
  &{DS vs K} &{-0.32} &{-} &{-0.66} &{-}\\
  \cline{2-6}
  &{ASA vs K} &{0.05} &{-} &{-0.10} &{-}\\
  \cline{2-6}
  &{SF vs K} &{0.67} &{-} &{-0.20} &{-}\\
 \hline
 \multicolumn{2}{c}{Number of clusters} &{4} &{5} &{3} &{3}\\
 \hline
 \multicolumn{2}{c}{Number of rays per cluster} &{3} &{5} &{3} &{2}\\
  \hline
 \multicolumn{2}{c}{Cluster DS ($\rm{C_{DS}}$) in [ns]} &{0.5} &{1.4} &{4.1} &{0.3}\\
 \hline
 \multicolumn{2}{c}{Cluster ASA ($\rm{C_{ASA}}$) in [$\degree$]} &{1.5} &{4.7} &{0.8} &{0.6}\\
 \hline
  \multicolumn{2}{c}{Cluster K-factor ($\rm{C_{K}}$) in [$\rm{dB}$]} &{1.47} &{-1.43} &{13.49} &{10.88}\\
 \hline
 \multirow{4}{*}{\makecell{Correlation distance in \\the horizontal plane [m]}} &{ASA} &{2.1} &{2.4} &{5.6} &{3.1}\\
  \cline{2-6}
  &{DS} &{1.9} &{1.0} &{4.9} &{4.7}\\
  \cline{2-6}
  &{SF} &{2} &{0.8} &{5.2} &{7.8}\\
  \cline{2-6}
  &{K} &{2.4} &{-} &{2.5} &{-}\\
  \hline
\end{tabular}
\end{center}
\end{table*}
\subsection{Cross-Correlations and correlation distance in the horizontal plane}\label{48}
The cross-correlations and the correlation distance in the horizontal plane of both scenarios are summarized in Table \ref{Table 2}.
For the indoor office, ASA shows a positive correlation with DS in both LoS and NLoS cases. This can be attributed to the dispersion of the channel in the angular domain, which is likely to be correlated with dispersion in the delay domain. The ASA and DS exhibit a positive correlation with SF in the LoS case, whereas they are negatively correlated with SF in the NLoS case. This is because SF means more reflection disperses the channel energy at both LoS and NLoS cases, but in the NLoS case, the MPCs are harder to be received. The DS is negatively correlated with the K-factor since a small K-factor means a weaker dominant LoS path and a more dispersive channel. The ASA does not show a correlation with the K-factor, indicating that the angular dispersion of MPCs in the office is sufficiently large and is not easily affected by the power of MPCs. The SF is negatively correlated to the K-factor because a larger SF causes the attenuation of the MPCs power, resulting in the dominant LoS path having a larger proportion.

The correlation distance in the horizontal plane is shown in Table \ref{Table 2}. For the indoor office, the correlation distance presents the maximum distance at which parameters of two location points are kept relevant. The correlation distance of DS and SF in the LoS case is larger than that in the NLoS case, which means DS and SF are more stationary over the distance in the LoS case. The correlation distance of ASA in the LoS case is smaller than that in the NLoS case, which indicates the MPCs propagating in a small and closed room (NLoS) exhibit greater spatial consistency in the angular domain.

For the UMi, the ASA and DS are positively correlated in the LoS case because both ASA and DS represent the dispersion of MPCs. Analogously, the ASA and DS are negatively correlated with the K-factor due to a small K-factor means the dominant LoS path is weaker and thus the channel is more dispersive. ASA and DS are negatively correlated in the NLoS case because a large DS means a long propagation distance and low receiving power. Thus, the MPCs scattered next to the first reflection direction have lower power, causing the power to be more concentrated in the first reflection direction. ASA and DS have a positive correlation with SF in the NLoS case. The larger SF in the NLoS case means more reflection, and MPCs with different DS and AoA. Besides, DS and ASA have a negative correlation with SF in the LoS case because more SF results in less power for MPCs. SF is negatively corrected to the K-factor because larger SF causes more MPCs by reflection and more MPCs can be received by the RX, thus the LoS path has a smaller proportion. Besides, it is found that the correlations between DS and SF, and between SF and K-factor for both two scenarios are quite different. The reason is that SF in the indoor office means attenuation of the power for all the MPCs while SF in the UMi means more reflections and MPCs have more chances to be received.

For the correlation distance in the UMi, the correlation distance of ASA and DS in the LoS case is larger than that in the NLoS case, indicating that these parameters exhibit greater spatial stationarity over the distance in the LoS case. In contrast, as a case with primary reflection as the main mode of propagation, the NLoS case has less variation in SF with distance. Besides, the correlation distance in the UMi is larger than that in the indoor office due to the wider range of distances present in the UMi scenario.

In the next section, for the observed THz channel sparsity in the propagation, delay, and spatial domains, we propose the THz channel model and modeling implementation framework to reconstruct the CIRs and verify the framework by the items of CIRs, Gini factors, and the channel capacity in the simulated THz channel.
\section{THZ CHANNEL MODEL and 
 SIMULATION IMPLEMENTATION FRAMEWORK}\label{5}
In this section, we investigate the THz 3GPP-like three-dimensional (3D) GBSM, propose a modified modeling implementation framework, and study the CIRs, Gini factors, and the channel capacity by the simulation of the channel modeling framework to verify the characterization of the THz channel sparsity in both the indoor office and UMi scenarios. 
\subsection{The principle of the 3GPP-like GBSM for THz bands}\label{34}
The 3GPP-like THz GBSM is introduced to describe the channel transfer function (CTF) in the THz channel in this subsection, which has the advantage of characterizing the characteristics of randomly distributed scatterers by the fundamental laws of electromagnetic wave reflection, diffraction, and scattering. The 3GPP-like THz GBSM models the rays by the 3D pattern of the antennas, and the arrival and departure angles to characterize the geometric feature of the channel. The CIF of the whole multiple-input multiple-output (MIMO) channel ${\bf{H}}(\tau,t)$ can be described by $M_\text{r} \times M_\text{t}$, where $M_\text{t}$ and $M_\text{r}$ are the number of elements at the TX and RX sides, respectively. The element of the matrix presented by the GBSM means the CTF for $s$th TX element and $u$th RX element and can be written as \cite{r10}:
\begin{equation}\label{GBSM}
\begin{split}
h_{u,s}(\tau,t)=&\sqrt{\frac{1}{K_\text{R}+1}}h^\text{NLoS}_{u,s}(\tau,t)+\\
&\sqrt{\frac{K_\text{R}}{K_\text{R}+1}}h^\text{LoS}_{u,s}(t)\delta(\tau-\tau^\text{LoS}),
\end{split}
\end{equation}
\noindent where $h^\text{NLoS}_{u,s}(\tau,t)$ is the CTF of clusters in the case of NLoS, $h^\text{LoS}_{u,s}(t)$ is the CTF of the LoS clusters, $\delta(\cdot)$ is the Dirac's delta function, and $\tau^\text{LoS}$ is the delay of the LoS cluster. Besides, $\tau_n$ denotes the delay of the $n$th cluster. $h^\text{NLoS}_{u,s}(\tau,t)$ can be expressed as \cite{r10}:
\begin{equation}\label{GBSMNLOS}
\begin{split}
h^\text{NLoS}_{u,s}(\tau,t)=&\sum\limits_{n=1}^2\sum\limits_{i=1}^3\sum\limits_{m\in R_i}h^\text{NLoS}_{u,s,n,m}(t)\delta(\tau-\tau_{n,i})+\\
&\sum\limits_{n=3}^Nh^\text{NLoS}_{u,s,n}(t)\delta(\tau-\tau_{n}),
\end{split}
\end{equation}
\noindent where $n$ is the cluster number, $N$ is the total cluster number, $m$ is the rays in the cluster, $R_i$ is the mapping to rays at sub-cluster $i$, $\tau_{n,i}$ is the delay of rays of sub-cluster $i$ at cluster $n$. The rays of the two strongest clusters ($n=1$ and $2$) are spread to delay to three sub-clusters (per cluster), with fixed delay offset.

It is worth noting that the measurement results for cluster number, cluster DS, and cluster ASA are much smaller than those in the 3GPP model, making the set of sub-clusters to compensate for delay offset unnecessary. The NLoS GBSM at THz can be simplified to:
\begin{equation}\label{GBSMNLOSS}
\begin{split}
h^\text{NLoS}_{u,s}(\tau,t)=\sum\limits_{n=1}^Nh^\text{NLoS}_{u,s,n}(t)\delta(\tau-\tau_{n}).
\end{split}
\end{equation}

To characterize the channel sparsity in the THz channel, the intra-cluster K-factors $K_\text{IC}$ are used to distribute the power in one cluster \cite{r37}. The CTF of the cluster $n$ $h^\text{NLoS}_{u,s,n}(t)$ is written as:
\begin{equation}\label{GBSMNLOSSS}
\begin{split}
h^\text{NLoS}_{u,s,n}(\tau,t)=&\sqrt{\frac{K_\text{IC}P_n}{1+K_\text{IC}}}\sum\limits_{m=1}^1h^\text{NLoS}_{u,s,n,m}(t)\\
&+\sqrt{\frac{P_n}{(1+K_\text{IC})(M-1)}}\sum\limits_{m=2}^Mh^\text{NLoS}_{u,s,n,m}(t),
\end{split}
\end{equation}

\begin{figure*}[ht]
\begin{equation}\label{GBSMNLOSMN}
\begin{split}
h^\text{NLoS}_{u,s,n,m}(t)=&
\begin{bmatrix}
F_{\text{rx},u,\theta}(\theta_{n,m,\text{ZoA}},\phi_{n,m,\text{AoA}})\\
F_{\text{rx},u,\phi}(\theta_{n,m,\text{ZoA}},\phi_{n,m,\text{AoA}})
\end{bmatrix}
^T
\begin{bmatrix}
{\rm{e}}^{{\rm{j}}\Phi_{n,m}^{\theta\theta}} & \sqrt{\kappa_{n,m}^{-1}}{\rm{e}}^{{\rm{j}}\Phi_{n,m}^{\theta\phi}}\\
\sqrt{\kappa_{n,m}^{1}}{\rm{e}}^{{\rm{j}}\Phi_{n,m}^{\phi\theta}} & {\rm{e}}^{{\rm{j}}\Phi_{n,m}^{\phi\phi}}
\end{bmatrix}
\begin{bmatrix}
F_{\text{tx},s,\theta}(\theta_{n,m,\text{ZoD}},\phi_{n,m,\text{AoD}})\\
F_{\text{tx},s,\phi}(\theta_{n,m,\text{ZoD}},\phi_{n,m,\text{AoD}})
\end{bmatrix}\\
&{\rm{e}}^{\frac{{\rm{j}}2\pi({\bf{r}}_{\text{rx},n,m}\cdot{\bf{d}}_{\text{rx},u})}{\lambda_0}}
{\rm{e}}^{\frac{{\rm{j}}2\pi({\bf{r}}_{\text{tx},n,m}\cdot{\bf{d}}_{\text{tx},s})}{\lambda_0}}{\rm{e}}^{\frac{{\rm{j}}2\pi({\bf{r}}_{\text{rx},n,m}\cdot{\bf{v}})t}{\lambda_0}}
\end{split}
\end{equation}
\hrulefill
\end{figure*}
\noindent and the CTF of the ray $m$ in the cluster $n$ $h^\text{NLoS}_{u,s,n,m}(t)$ is written as Eq. (\ref{GBSMNLOSMN}), where
\begin{itemize}
\item $[\cdot]^T$ is the transposition of the matrix.
\item $M$ is the number of the total rays in a cluster.
\item $P_n$ denotes the power of the cluster $n$.
\item $F_{\text{rx},u,\theta}$, $F_{\text{rx},u,\phi}$, $F_{\text{tx},s,\theta}$, $F_{\text{tx},s,\phi}$ denote the radiation pattern of elements $u$ at RX of the vertical and horizontal polarization and elements $s$ at TX of the vertical and horizontal polarization, respectively. Besides, $\theta$ represents the vertical polarization and $\phi$ represents the horizontal polarization.
\item $\theta_{n,m,\text{ZoA}}$, $\phi_{n,m,\text{AoA}}$, $\theta_{n,m,\text{ZoD}}$, and $\phi_{n,m,\text{AoD}}$ are the zenith angle of arrival (ZoA), the azimuth angle of arrival (AoA), the zenith angle of departure (ZoD), and the azimuth angle of departure (AoD) of the $m$th ray in $n$th cluster.
\item $\Phi_{n,m}^{\theta\theta}$, $\Phi_{n,m}^{\theta\phi}$, $\Phi_{n,m}^{\phi\theta}$, and $\Phi_{n,m}^{\phi\phi}$ are the random phase for four different polarization combinations ($\theta\theta$, $\theta\phi$, $\phi\theta$, $\phi\phi$) of the $m$th ray in $n$th cluster. Besides, $\kappa_{n,m}$ is the cross-polarization power ratio (XPR) of the $m$th ray in the $n$th cluster.
\item ${\bf{r}}_{\text{rx},n,m}$ and ${\bf{r}}_{\text{tx},n,m}$ denotes the spherical unit vector at RX with $\theta_{n,m,\text{ZoA}}$ and $\phi_{n,m,\text{AoA}}$ and at TX with $\theta_{n,m,\text{ZoD}}$ and $\phi_{n,m,\text{AoD}}$, respectively. For example, ${\bf{r}}_{\text{rx},n,m}$ is given by:
\begin{equation}\label{vector}
\begin{split}
\begin{bmatrix}
\sin{\theta_{n,nm,\text{ZoA}}}\cos{\phi_{n,m,\text{AoA}}}\\
\sin{\theta_{n,nm,\text{ZoA}}}\sin{\phi_{n,m,\text{AoA}}}\\
\cos{\theta_{n,m,\text{ZoA}}}
\end{bmatrix}^{T}.
\end{split}
\end{equation}
\item ${\bf{d}}_{\text{rx},u}$ and ${\bf{d}}_{\text{tx},s}$ denotes the location vector of receive antenna element $u$ and transmit antenna element $s$. 
\item $\frac{{\bf{r}}_{\text{rx},n,m}\cdot{\bf{v}}}{\lambda_0}$ is the Doppler frequency component. $\lambda_0$ is the wavelength of the carrier frequency. The ${\bf{v}}$ is written as:
\begin{equation}\label{vvector}
\begin{split}
{\bf{v}}=v\cdot
\begin{bmatrix}
\sin{\theta_v}\cos{\phi_v} & \sin{\theta_v}\sin{\phi_v} & \cos{\theta_v}
\end{bmatrix}
^T,
\end{split}
\end{equation}
\begin{figure*}[ht]
\begin{equation}\label{GBSMLOS}
\begin{split}
h^\text{LoS}_{u,s}(t)=
&\begin{bmatrix}
F_{\text{rx},u,\theta}(\theta_{\text{LoS},\text{ZoA}},\phi_{\text{LoS},\text{AoA}})\\
F_{\text{rx},u,\phi}(\theta_{\text{LoS},\text{ZoA}},\phi_{\text{LoS},\text{AoA}})
\end{bmatrix}
^T
\begin{bmatrix}
1 & 0\\
0 & -1
\end{bmatrix}
\begin{bmatrix}
F_{\text{tx},s,\theta}(\theta_{\text{LoS},\text{ZoD}},\phi_{\text{LoS},\text{AoD}})\\
F_{\text{tx},s,\phi}(\theta_{\text{LoS},\text{ZoD}},\phi_{\text{LoS},\text{AoD}})
\end{bmatrix}
{\rm{e}}^{-{\rm{j}}2\pi\frac{d}{\lambda_0}}
{\rm{e}}^{\frac{{\rm{j}}2\pi({\bf{r}}_{\text{rx},\text{LoS}}\cdot{\bf{d}}_{\text{rx},u})}{\lambda_0}}\\
&{\rm{e}}^{\frac{{\rm{j}}2\pi({\bf{r}}_{\text{tx},\text{LoS}}\cdot{\bf{d}}_{\text{tx},s})}{\lambda_0}}{\rm{e}}^{\frac{{\rm{j}}2\pi({\bf{r}}_{\text{rx},\text{LoS}}\cdot{\bf{v}})t}{\lambda_0}}
\end{split}
\end{equation}
\hrulefill
\end{figure*}where $v$ is the user velocity vector, $\theta_v$ is the travel elevation angle, and $\phi_v$ is the travel azimuth angle.
\end{itemize}

In the LoS case, the $h^\text{LoS}_{u,s}(t)$ can be expressed as Eq. (\ref{GBSMLOS}), where $-2\pi\frac{d}{\lambda_0}$ is the initial phase for both $\theta\theta$ and $\phi\phi$ polarization of the LoS link. 
\subsection{The simulation implementation framework of the proposed THz GBSM}\label{344}
The THz channel exhibits distinct characteristics due to the unique properties of THz waves, such as their short wavelength. For instance, the composition of clusters, which serves as the fundamental unit in the GBSM, may undergo changes. Therefore, except for the modifications in the THz model, there are several revisions for the traditional 3GPP GBSM implementation framework to model various types of the THz channel parameters:
\begin{itemize}
\item The THz spread parameters, cross-correlations, and correlation distance of the spread parameters are utilized to generate correlated large-scale parameters, namely DS, angle spread (AS), shadow fading, and K-factor. The large-scale parameters are represented by the lognormal distribution, as summarized in Table \ref{Table 2} in both scenarios. The small-scale parameters (delay, AoA, AoD, ZoA, ZoD) in the GBSM are computed based on the correlated large-scale parameters. 
\item The powers of the paths in each cluster and intra-cluster are calculated by the THz PL, K-factor, and cluster K-factor \cite{r37}. The PL is modeled by the CI model.
\item The cluster parameters such as the THz cluster number, number of rays per cluster, in-cluster DS, and in-cluster AS are utilized to divide the MPCs into clusters to describe the CTF of the channel.
\end{itemize}

\begin{figure}[!t]
\centerline{\includegraphics[width=8.5 cm, trim=2cm 1.5cm 2cm 1.5cm, clip]{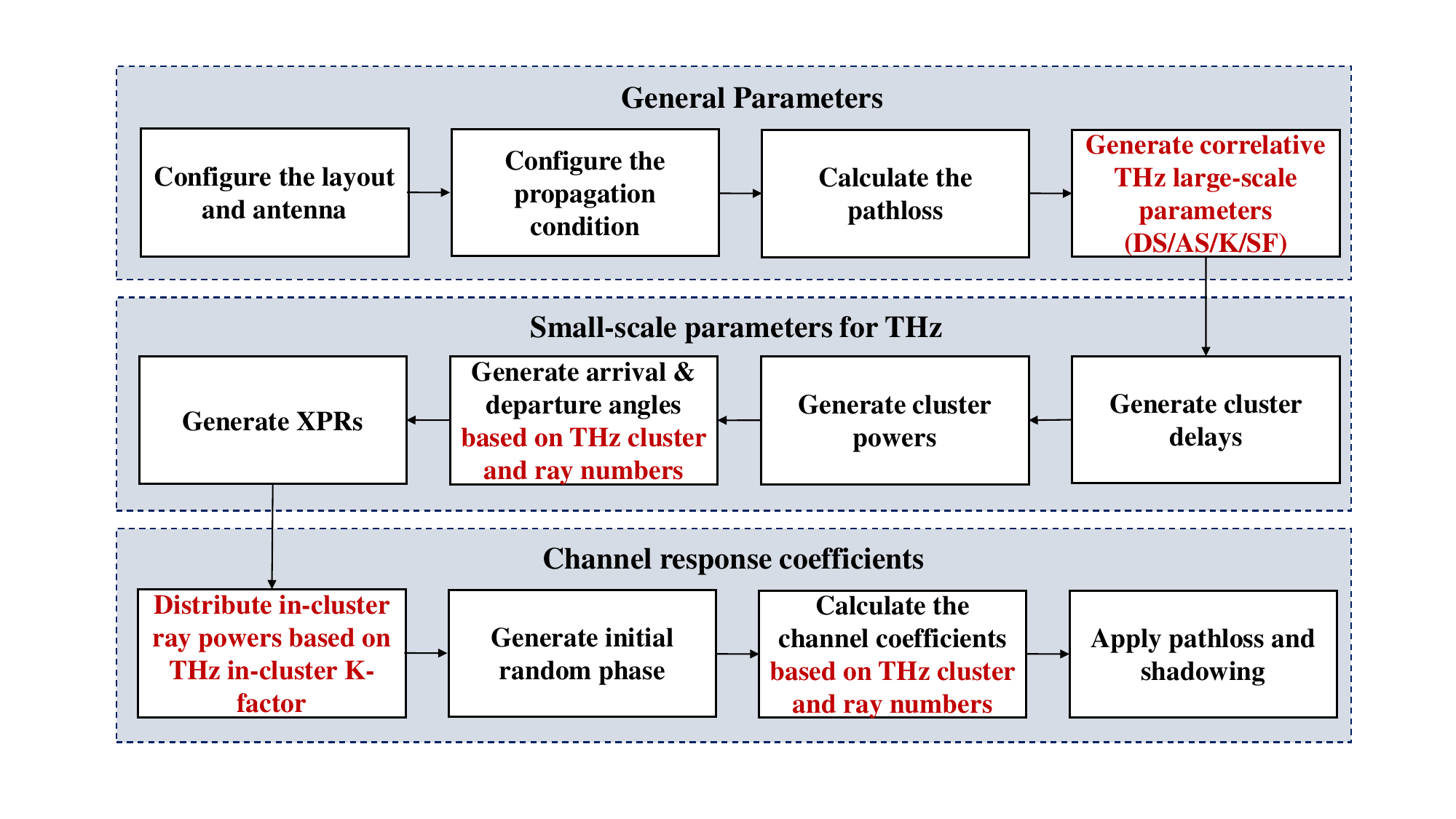}}
\caption{THz channel 3GPP-like GBSM implementation framework.}\label{Fig GBSM}
\end{figure}

To sum up, the THz GBSM implementation framework is shown in Fig. \ref{Fig GBSM}, which is an extension of the current 3GPP GBSM framework. The modifications are noted in red fonts. These include generating correlated large-scale parameters in the THz bands based on Table. \ref{Table 2}, generating arrival and departure angles based on the THz large-scale parameters, distributing cluster powers and intra-cluster ray powers by THz K-factor and intra-cluster K-factor \cite{r37}, respectively, and calculating the channel coefficients based on the cluster and ray numbers at THz. All the required parameters of modeling are shown in Table \ref{Table 2}. The simulation of the THz GBSM is conducted using the proposed implementation framework and simulation software \cite{r33}, following a step-by-step approach \cite{r10}. 
\begin{table}[t]
\begin{center}
\setlength{\tabcolsep}{12pt}  
\footnotesize
\renewcommand\arraystretch{1.3}  
\caption{Simulation Parameters of THz channel modeling framework.}\label{Table 3}
\vglue8pt
\begin{tabular}{ccc}  
 \hline
   {\textbf{Scenario}}  &{\textbf{Indoor office}} & {\textbf{UMi}}\\
    \hline
       {Carrier frequency}  &{100 $\rm{GHz}$} &{132 $\rm{GHz}$}\\
    \hline
   {BS height}  &{3 $\rm{m}$} & {10 $\rm{m}$}\\
    \hline 
   {MUs height}   &{1.5 $\rm{m}$} & {1.5 $\rm{m}$}\\
    \hline
   {Bandwidth}  &{1.2 $\rm{GHz}$}  &{1.2 $\rm{GHz}$}\\
    \hline
   {BS transmit power}  &{35 $\rm{dBm}$} &{35 $\rm{dBm}$}\\
    \hline
   {BS antenna gain}   &{5 $\rm{dBi}$} &{5 $\rm{dBi}$}\\
    \hline
   {MUs antenna gain}  &{5 $\rm{dBi}$} &{5 $\rm{dBi}$}\\
    \hline
\end{tabular}
\end{center}
\end{table}
\subsection{The verification of the THz channel model}\label{3444}
To verify the performance of the proposed THz channel model and its implementation framework, we simulate the THz channel for two scenarios and evaluate the channel capacity. The details of parameters are shown in Table \ref{Table 3}.

\begin{figure}[!t]
\centering
\subfigure[Indoor office]{\includegraphics[width=7cm]{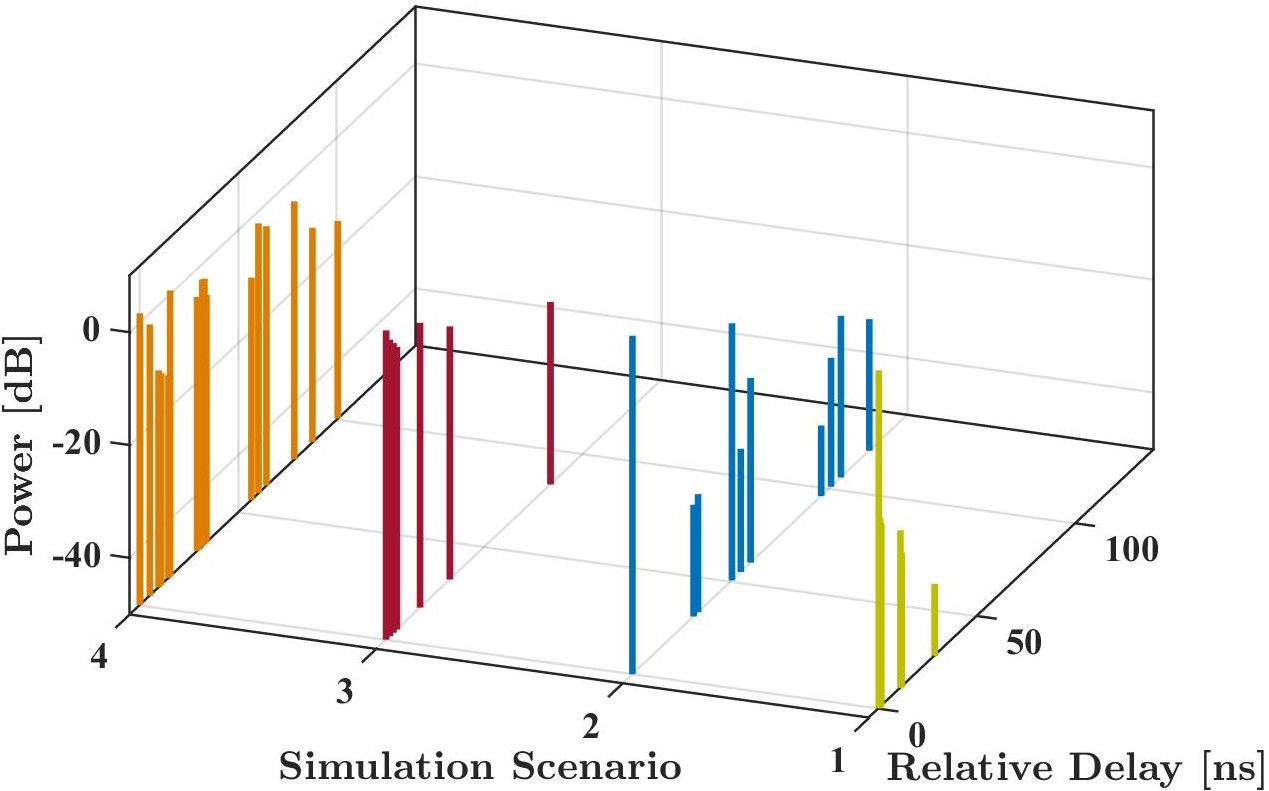}}\\
\subfigure[UMi]{\includegraphics[width=7cm]{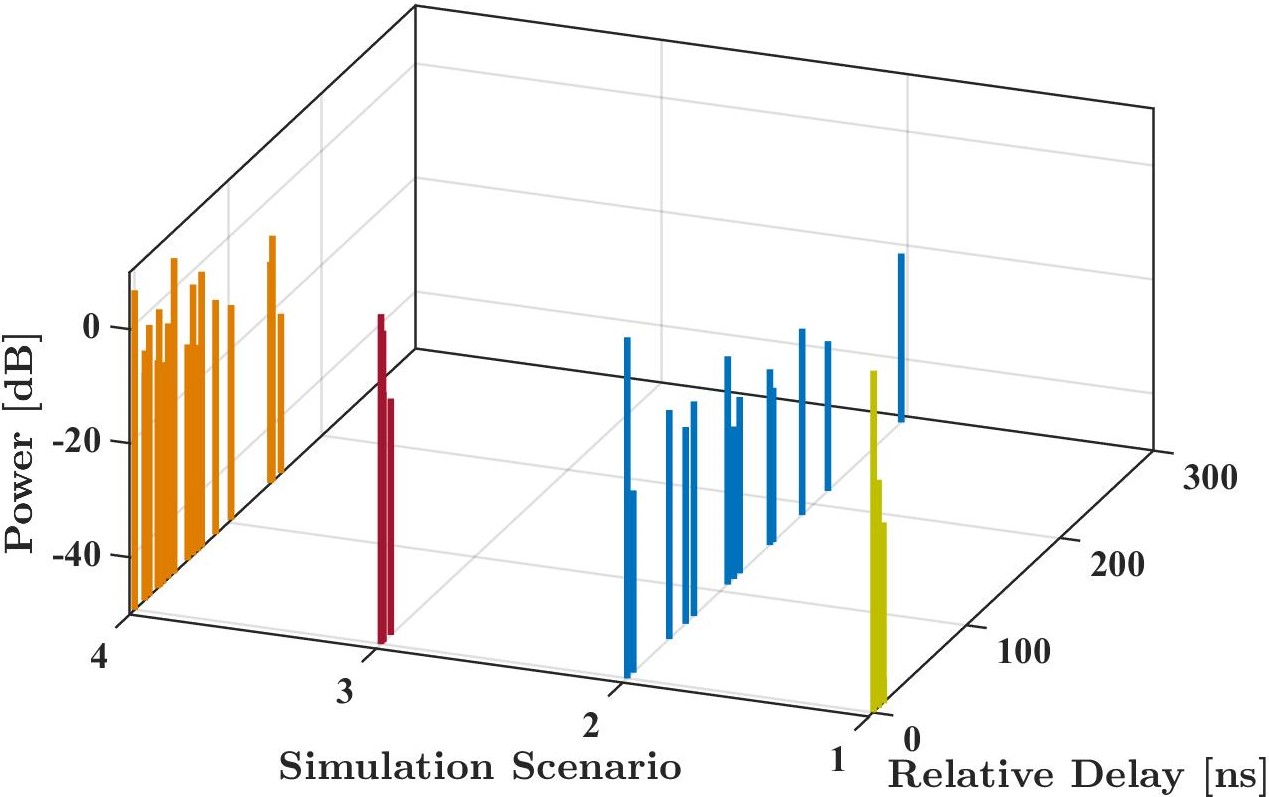}}
\caption{The simulated result of CIRs at 100 GHz in the indoor office and at 132 GHz in the UMi.}\label{Fig 15}
\end{figure}

The simulated communication links are operating at 100 $\rm{GHz}$ in the indoor office and 132 $\rm{GHz}$ in the UMi scenario by the proposed THz channel modeling framework and the 3GPP model. Fig. \ref{Fig 15} shows the simulation CIRs based on clusters using the proposed framework. As a reference, the lower frequency 3GPP model is shown in the Fig. \ref{Fig 15}. Scenario indexes 1 and 3 denote simulation results by the proposed THz channel modeling framework in the LoS and NLoS case, respectively, and indexes 2 and 4 are the scenarios simulated by the 3GPP model for reference. It is observed that the THz channel model can describe the sparse THz channel better than the reference 3GPP model by the specific cluster numbers, and the inter-cluster and intra-cluster K-factors. The indoor office appears more dispersed clusters in the delay domain than the UMi, which can also be noticed in the measurement results. Besides, the UMi results in Fig. \ref{Fig 15}b also describe the street canyon scenarios well, especially in the NLoS case with few clusters that have similar delays. To further investigate the accuracy of the large-scale parameters for the obtained CIRs, 10000 times simulations are conducted for the random layout of the UEs, and the large-scale parameters are shown in Table \ref{Table 6}. The parameters present accurate results with a max deviation of about 0.2, as the extensive simulation times and the single antenna in the ports of BS and UEs are set in the simulation.

\begin{figure}[!t]
\centering
\subfigure[LoS]{\includegraphics[width=6.5cm]{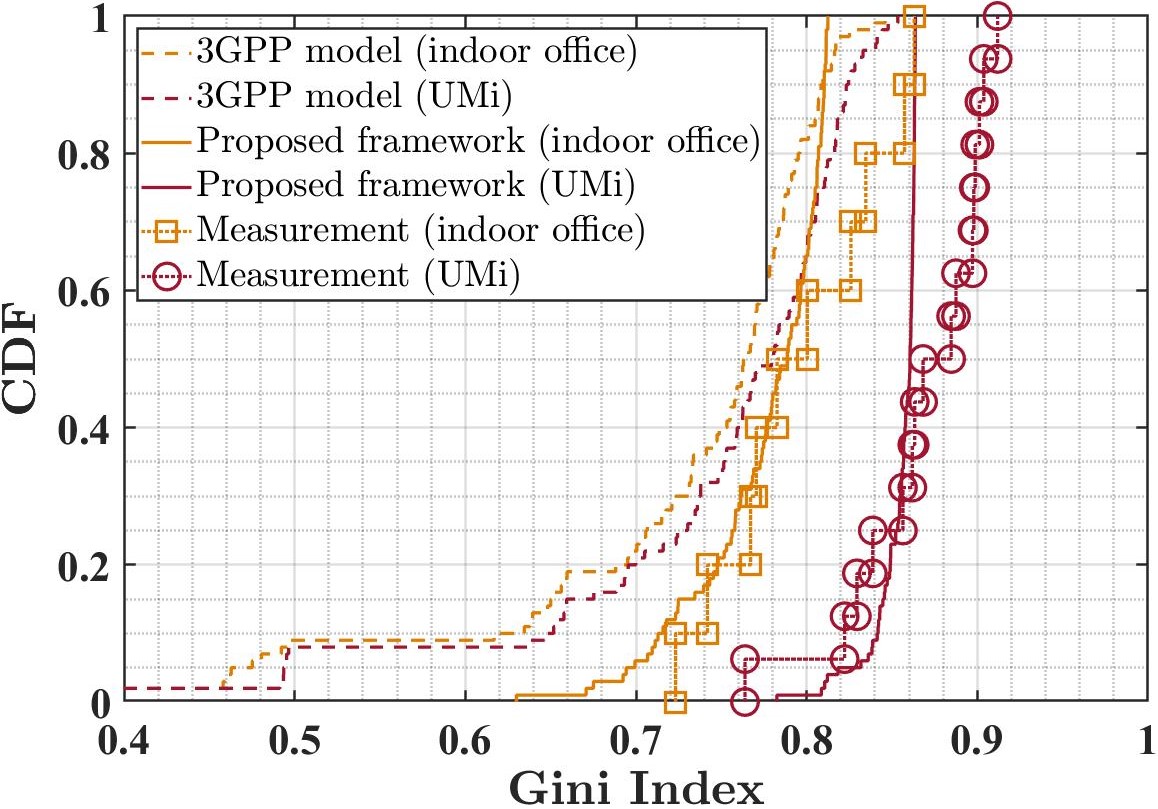}}\\
\subfigure[NLoS]{\includegraphics[width=6.5cm]{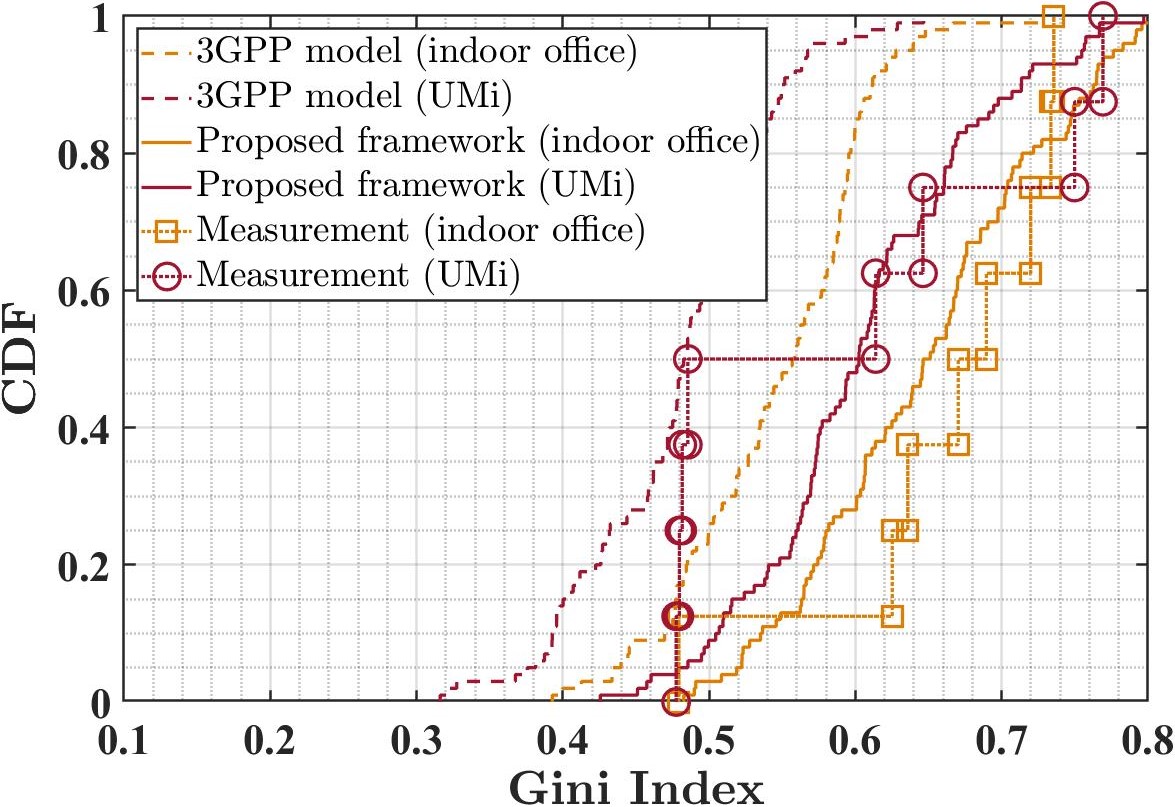}}
\caption{The simulated result of Gini factor at 100 GHz in the indoor office and at 132 GHz in the UMi.}\label{Fig 16}
\end{figure}

\begin{figure}[!t]
\centering
\centerline{\includegraphics[width=6.45cm]{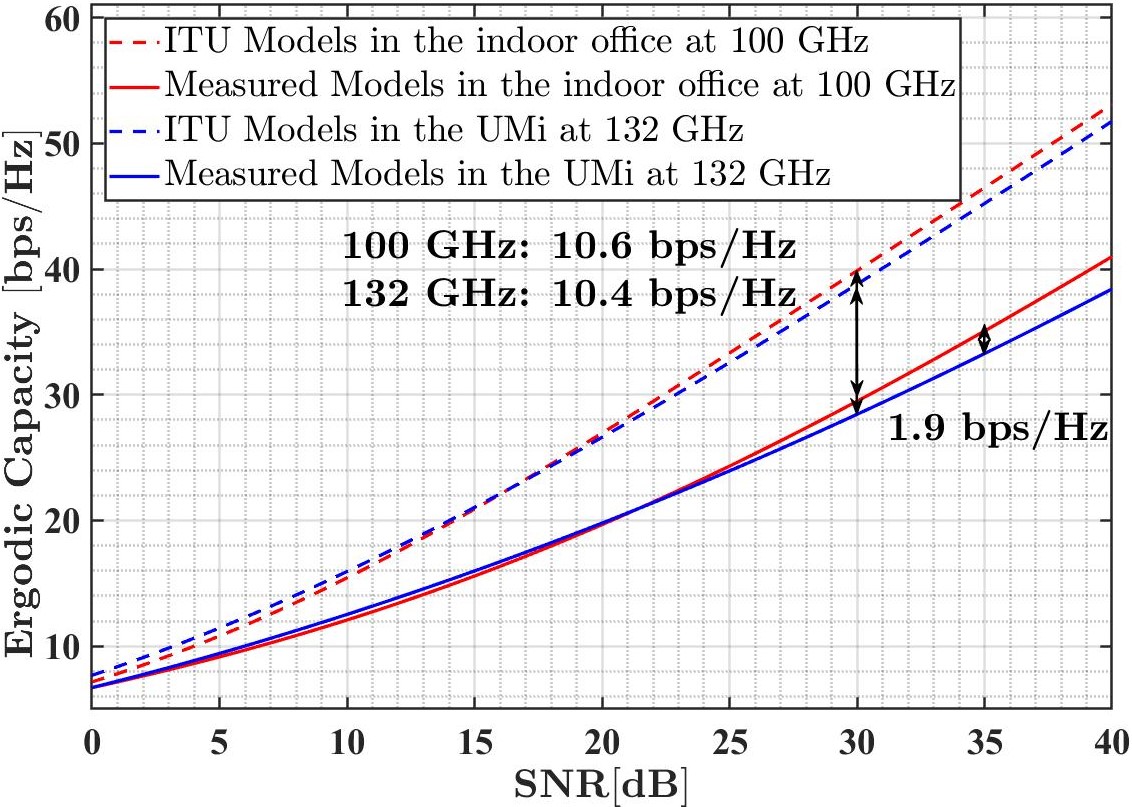}}
\caption{The channel capacity for measurement and 3GPP model at 100 $\rm{GHz}$ in the indoor office and at 132 $\rm{GHz}$ in the UMi.}\label{Fig 13}
\end{figure}

As the simulation results show, the channel sparsity has been described by the proposed THz channel modeling framework. To further evaluate the correctness of the channel sparsity characterizing, the Gini factor is utilized which has been introduced in \cite{r37}. Fig. \ref{Fig 16} shows the CDF of the Gini factors calculated by the CIRs, which is obtained by the 3GPP model, proposed THz channel modeling framework, and channel measurements. 100 samples are used in the simulation. In the LoS case, the 3GPP model results present a long tail and similar CDF in both scenarios with the minimum error of 0.04 and 0.1 in the indoor and outdoor scenarios, while the proposed THz channel model gives a maximum deviation of 0.04 in both scenarios and conducts a good fitting results at most cumulative probability. In the NLoS case, although the proposed model shows a large deviation in the low cumulative probability in the UMi scenario as its much sparse measurements results, the model also improves the characterizing ability of the channel sparsity in the items of trend and correctness. 
\begin{table}[t]
\begin{center}
\setlength{\tabcolsep}{8pt}  
\footnotesize
\renewcommand\arraystretch{1.3}  
\caption{Obtained large scale parameters using the THz channel modeling framework.}\label{Table 6}
\vglue8pt
 \begin{tabular}{ccccc}  
 \hline
  {\textbf{Parameter}} &{\textbf{Scenario}}  &{\textbf{Mean $\mu$}} &{\textbf{Sigma $\sigma$}} &{\textbf{PLE}} \\     
 \hline
 \multirow{4}{*}{RMS DS} &{Indoor LoS} &{-8.82} &{0.15} &{-}\\
  \cline{2-5}
  &{Indoor NLoS} &{-8.10} &{0.15} &{-}  \\
  \cline{2-5}
  &{UMi LoS} &{-8.19} &{0.56} &{-}  \\
  \cline{2-5}
  &{UMi NLoS} &{-8.52} &{0.17} &{-}  \\
  \hline
  \multirow{4}{*}{RMS ASA} &{Indoor LoS} &{1.36} &{0.21} &{-} \\
  \cline{2-5}
  &{Indoor NLoS} &{1.62} &{0.1} &{-}  \\
  \cline{2-5}
  &{UMi LoS} &{1.36} &{0.21} &{-} \\
  \cline{2-5}
  &{UMi NLoS} &{1.62} &{0.1} &{-}\\
  \hline
  \multirow{4}{*}{K-factor} &{Indoor LoS} &{8.58} &{5.04} &{-}\\
  \cline{2-5}
  &{Indoor NLoS} &{-} &{-} &{-}\\
  \cline{2-5}
  &{UMi LoS} &{18.78} &{6.28} &{-}\\
  \cline{2-5}
  &{UMi NLoS} &{-} &{-} &{-}\\
  \hline
  \multirow{4}{*}{Path loss} &{Indoor LoS} &{-} &{-} &{1.94}\\
  \cline{2-5}
  &{Indoor NLoS} &{-} &{-} &{2.78}\\
  \cline{2-5}
  &{UMi LoS} &{-} &{-} &{1.98}\\
  \cline{2-5}
  &{UMi NLoS} &{-} &{-} &{2.50}\\
  \hline
\end{tabular}
\end{center}
\end{table}
Finally, to investigate the performance of the THz communication system, we evaluate the channel capacity of THz communication systems in both the indoor office and UMi scenarios by the proposed framework, and the 3GPP model is used for reference. The simulated communication links are operating at 100 $\rm{GHz}$ in the indoor office and 132 $\rm{GHz}$ in the UMi scenario with 16 $\times$ 16 omnidirectional antennas BS and 2 $\times$ 2 omnidirectional antennas MUs. The numbers of BS and MUs antenna refer to \cite{r26}. The details of parameters are shown in Table \ref{Table 3}. The channel capacity defined as the maximal achievable transmission rate is calculated by:
\begin{equation}\label{cap}
\begin{split}
C=\log_{2}{\rm{det}}[{\bf{I}}_{M_r}+\frac{\rho}{M_t}{\bf{H}}(\tau,t){\bf{H}}^H(\tau,t)],
\end{split}
\end{equation}
\noindent where $M_r$ and $M_t$ are the numbers of MUs and BS antenna, which are 4 and 256, respectively. Besides, $\rm{det}$ presents computing the value of the determinant of the matrix, ${\bf{I}}_{M_r}$ is a unit matrix with the dimension of $M_r \times M_r$, $\rho$ is the mean SNR, ${{\bf{H}}(\tau,t)}$ is the $M_r \times M_t$ MIMO channel matrix which is the simulation result, and superscript ${H}$ represents the conjugate transpose of a matrix. Fig. \ref{Fig 13} presents the channel capacity obtained by the 3GPP Indoor Hotspot scenario model at 100 $\rm{GHz}$ \cite{r10}, 3GPP UMi scenario model at 132 $\rm{GHz}$, and obtained 100 $\rm{GHz}$ and 132 $\rm{GHz}$ channel parameters. The parameters not obtained by the measurement at 100 $\rm{GHz}$ and 132 $\rm{GHz}$ are substituted with corresponding 3GPP parameters. Besides, the outdoor to indoor probability is set to 0 in both scenarios. When the SNR exceeds 21.6 $\rm{dB}$, the channel capacity evaluated by the measurement model in the indoor scenario is greater than that in the UMi scenario (equivalent to a 1.9 $\rm{bps/Hz}$ gain at the SNR of 35 $\rm{dB}$). This is because the number of clusters in the UMi scenario is lower than that in the indoor office for both LoS and NLoS cases. Furthermore, it has been observed that the proposed THz model predicts lower performance when compared to the 3GPP model, both for the 100 $\rm{GHz}$ frequency band in the indoor office and the 132 $\rm{GHz}$ frequency band in the UMi scenario (equivalent to a 10.6 $\rm{bps/Hz}$ loss at 100 GHz in the indoor office and a 10.4 $\rm{bps/Hz}$ loss at 132 GHz in the UMi at the SNR of 30 $\rm{dB}$). This is because the THz channel contains little clusters and rays, which can be seen in Fig. \ref{Fig 15} (e.g., at 100 $\rm{GHz}$, the 3GPP model predicted 15 clusters while 4 clusters were measured in the indoor-hotspot scenario in the LoS case) and the proposed THz channel model takes into account the sparsity of the channel by updating the THz characteristic parameters and redistributing the power of the in-cluster rays. Alerted by the low capacity, in order to obtain better system performance, it is crucial to leverage the large bandwidth characteristics of THz in the design of THz communication systems. 
\section{CONCLUSION}\label{6}
This paper focuses on the 3GPP-like GBSM THz channel modeling based on the measurements at 100 $\rm{GHz}$ in the indoor office and at 132 $\rm{GHz}$ in the UMi scenario. Then, the 3GPP-like channel statistical models of characteristic parameters are extracted, and analyzed referred to the default models in the 3GPP channel model \cite{r10}. Through the analysis of the values, we find that the UMi exhibits a sparser MPC structure compared to the indoor office scenario. Specifically, the ASA for LoS and NLoS cases in the UMi is 1.13 $\log_{10}(\degree)$ and 0.59 $\log_{10}(\degree)$, respectively, which is much smaller than 1.37 $\log_{10}(\degree)$ and 1.62 $\log_{10}(\degree)$ in the office scenario. Besides, it is observed that the cluster numbers are 4 in the indoor office and 3 in the UMi in the LoS case, which are less than those in the 3GPP of 15 and 12, respectively. It means that the THz channel is more sparse and the existing channel models are unavailable to characterize the sparsity above 100 $\rm{GHz}$. Furthermore, a large-scale parameters table is presented and these parameters can be used in the channel model standardization for 6G. Finally, we propose the THz channel modeling framework to reconstruct CIRs to characterize the sparser THz channel based on the obtained models. The obvious channel sparsity is characterized in both scenarios, as the Gini factors obtained by the proposed model only have the maximum deviation of 0.04 for those of the measurement, and the large-scale parameters are characterized by the proposed model accurately. Moreover, the channel capacity of the THz channel is evaluated by the 3GPP and proposed model. The proposed THz modeling framework leads to a lower capacity (equivalent to more than 10 $\rm{bps/Hz}$ at the signal-to-noise ratio of 30 $\rm{dB}$). Overall, this work contributes to the modeling of the THz channels, the standardization of the THz channel models for 6G, and provides insights for the design and optimization of future 6G communication systems.

As for future work, there are two directions: i) Conduct additional measurements for both scenarios, particularly for the NLoS case, to validate the model in terms of data volume; ii) Investigate the impact of various environmental factors on the THz NLoS channel, such as buildings and echo reflector. 
\bibliographystyle{IEEEtran}
\bibliography{bib}
\end{document}